\documentclass[12pt,twoside]{article}
\usepackage{correl12}
\usepackage[innercaption]{sidecap}

\def\TheTitle{The Gutzwiller Density Functional Theory}
\def\TheHeading{Gutzwiller Theory}                        
\def\TheAuthor{J\"org B\"unemann}                              
\def\TheInstitute{Computational Physics Group}  
\def\TheAddress{BTU Cottbus}         
\def\TheChapter{5} 
   \vspace{-2cm}                                   

\newcommand{\ve}[1]{{\bm #1 }}
\newcommand{\veck}{{\bf k}}
\newcommand{\hcd}{\hat{c}^{\dagger}}
\newcommand{\hc}{\hat{c}^{\phantom{\dagger}}}
\newcommand{\ket}[1]{\left| #1 \right \rangle }  
\newcommand{\bra}[1]{\left \langle #1 \right |} 
\newcommand{\hh}{\hat{h}^{\phantom{\dagger}}}
\newcommand{\hhd}{\hat{h}^{\dagger}}
\begin{document}
\MakeTitle           
\tableofcontents     
\newpage
\section{Introduction}

\subsection{The electronic many-particle problem}
The fundamental Hamiltonian for the electrons in solid-state theory
has the well-known form 
($\hbar\equiv 1$)
\begin{eqnarray}
\label{1}
\hat{H}_{\rm el}&=&\sum_s\int {\rm d}^3r\; \hat{\psi}^{\dagger}_{s}(\ve{r})
\left(
-\frac{\Delta_{\ve{r}}}{2m}+V(\ve{r})
\right)
\hat{\psi}^{}_{s}(\ve{r})\\[3pt]
\nonumber
&&+\frac{1}{2}\sum_{s,s'}\int {\rm d}^3r\int {\rm d}^3r'\;
\hat{\psi}^{\dagger}_{s}(\ve{r})\hat{\psi}^{\dagger}_{s'}(\ve{r}')
\frac{e^2}{|\ve{r}-\ve{r}'|}\hat{\psi}^{}_{s'}(\ve{r}')
\hat{\psi}^{}_{s}(\ve{r})\;.
\end{eqnarray}
In these lecture notes, we assume that the potential $V(\ve{r})$, 
generated by the atomic nuclei, is perfectly lattice periodic. 
The operators 
$\hat{\psi}^{(\dagger)}_{s}(\ve{r})$ annihilate (create) an 
electron at the real-space position~$\ve{r}$ with spin $s=\uparrow,\downarrow$.
Despite the fact that the Hamiltonian~(\ref{1})
only describes the electronic degrees of freedom, the calculation
of the electrons' properties
poses a difficult many-particle problem which cannot be solved in general.
The strategies to deal with the many-body problem~(\ref{1}) 
can be grouped into two main categories.
\begin{itemize}
\item[i)]{\sl Model-system approaches}: 

In order to explain experiments, 
it is often sufficient to take into account only a limited number 
of degrees of freedom in the Hamiltonian~(\ref{1}). Therefore, the full problem
is replaced by a simpler `model Hamiltonian' which  
describes certain electronic properties of a material.
Celebrated examples are the Heisenberg model 
for magnetic insulators 
and the BCS Hamiltonian for superconductors.
In many-particle theory, in general, and for 
transition metals and their compounds, in particular, 
multi-band Hubbard Hamiltonians provide 
the standard models, see Sec.~\ref{sec3}. 

\item[ii)]{\sl Ab-initio approaches}: 

In order to cope with the full Hamiltonian~(\ref{1}), 
one has to resort to approximations
which are necessarily cruder than those designed 
for the investigation of much simpler model systems. The 
most frequently used {\sl ab-initio\/} approach is the 
`Local-Density Approximation' (LDA) 
 to `Density-Functional Theory' (DFT), 
see Sec.~\ref{sec2}.
\end{itemize}

\subsection{Density-Functional Theory (DFT)}\label{sec2}

The `Density-Functional Theory' (DFT) is based on a 
theorem by Hohenberg and Kohn~\cite{hohenberg1964}.
It states that there exists a universal functional 
$W\!\left\{n(\ve{r})\right\}$
 of the electronic density $n(\ve{r})$ such that
\begin{equation}
E\!\left\{n(\ve{r})\right\}=
\int {\rm d}^3r V(\ve{r})n(\ve{r})+W\!\left\{n(\ve{r})\right\}
\end{equation}  
has its minimum, $E_{0}\equiv E\!\left\{n_0(\ve{r})\right\}$, 
at the exact ground-state 
density $n_0(\ve{r})$ of the Hamiltonian~(\ref{1}) and $E_{0}$ 
is the corresponding ground-state energy. 
Since it is impossible to determine the functional $W\!\left\{n(\ve{r})\right\}$ 
for many-particle systems exactly, it is necessary 
to develop reasonable approximations for it.  
Usually, one writes
\begin{equation}
\label{2.260}
W\!\left\{n(\ve{r})\right\}=T\!\left\{n(\ve{r})\right\}+\frac{e^2}{2}
\int {\rm d}^3r\int {\rm d}^3r'\;
\frac{n(\ve{r})n(\ve{r}') }{|\ve{r}-\ve{r}'|}
+E_{\rm xc}\!\left\{n(\ve{r})\right\}\;,
\end{equation} 
where $T\!\left\{n(\ve{r})\right\}$ is the `kinetic-energy functional'
and $E_{\rm xc}\!\left\{n(\ve{r})\right\}$ 
is the `exchange-correlation functional' 
which contains all Coulomb-energy contributions 
apart from the Hartree term that was separated in~(\ref{2.260}).
Both functionals are unknown.
Approximate expressions for 
$T\!\left\{n(\ve{r})\right\}$ and $E_{\rm xc}\!\left\{n(\ve{r})\right\}$ 
are usually derived by considering a free electron gas. 
The kinetic energy of such electrons   
in the Hartree--Fock approximation is $\sim n^{5/3}$ 
where~$n$ is the density of the homogeneous electron gas.
Therefore, a common approximation for the kinetic-energy functional
in~(\ref{2.260}) is 
\begin{equation}
\label{2.270}
T\!\left\{n(\ve{r})\right\}=\frac{3}{10m}(3\pi^2)^{2/3}
\int {\rm d}^3r \;n(\ve{r})^{5/3}\;.
\end{equation} 
In the same way, one may approximate the exchange-correlation 
potential in the form
\begin{equation}
\label{2.280}
E_{\rm xc}\!\left\{n(\ve{r})\right\}
=- \int {\rm d}^3r \frac{3 e^2}{4\pi}(3\pi)^{1/3}n(\ve{r})^{4/3}\;.
\end{equation} 
To work with the energy functionals~(\ref{2.270}) 
and~(\ref{2.280}) is a simple example of a `Local-Density 
Approximation' (LDA) because only the local density appears
in $W\!\left\{n(\ve{r})\right\}$ and corrections, e.g., involving
gradients $\nabla n(\ve{r})$, are absent.

The DFT in LDA, as introduced so far, provides an approximate 
way to determine the ground-state energy and the electronic 
density in the ground state. These quantities are of interest
if one aims, e.g., to determine 
the ground-state lattice structure or lattice parameters of a material. 
Most practical applications of the DFT, however, rely on an additional 
concept, the `Kohn--Sham scheme'. 
This scheme is based on the 
assumption that, for each system of interacting particles, there 
exists an effective single-particle Hamiltonian 
\begin{eqnarray}
\label{2.290}
\hat{H}^{\rm eff}_0&=&\sum_s\int {\rm d}^3r\hat{\psi}^{\dagger}_{s}(\ve{r})
\left[-\frac{\Delta_{\ve{r}}}{2m}
+V(\ve{r})
 \right]\hat{\psi}^{}_{s}(\ve{r})
\\[3pt]
\nonumber
&&+\sum_s\int  {\rm d}^3r\hat{\psi}^{\dagger}_{s}(\ve{r})
\left[
e^2\int {\rm d}^3r'\frac{n(\ve{r}')}{|\ve{r}-\ve{r}'|}
+V^{\rm KS}_{\rm xc}\!\left\{n(\ve{r})\right\}
\right] \hat{\psi}^{}_{s}(\ve{r})\;,
\end{eqnarray}
which has the same ground-state density $n_0(\ve{r})$ 
as the original many-particle Hamiltonian~(\ref{1}).
In general, one cannot prove rigorously that such 
a single-particle Hamiltonian exists; this poses the
`$v$-representability' problem. If a system is $v$-representable,
however, the `Kohn--Sham potential' in~(\ref{2.290}) 
is given by
\begin{equation} 
\label{2.291}
V^{\rm KS}_{\rm xc}\!\left\{n(\ve{r})\right\}=
\left.\frac{\partial 
}{\partial \tilde{n}(\ve{r})}
\biggl(T\left\{\tilde{n}(\ve{r})\right\}-
T'\left\{\tilde{n}(\ve{r})\right\}
+E_{\rm xc}\left\{\tilde{n}(\ve{r})\right\}\biggr)
\right|_{\tilde{n}(\ve{r})=n(\ve{r})}\;,
\end{equation}
where $T'\!\left\{n(\ve{r})\right\}$ is the minimum kinetic 
energy of free non-interacting 
particles with the density distribution $n(\ve{r})$. 
Usually one assumes $T'\!\left\{n(\ve{r})\right\}=T\!\left\{n(\ve{r})\right\}$ 
such that both terms cancel each other in~(\ref{2.291}). 
 
For our formulation of a self-consistent `Gutzwiller Density-Functional 
Theory` (GDFT) in Sec.~\ref{gdft},
it will be convenient to introduce a basis of local (`Wannier') orbitals 
$\phi_{i,\sigma}(\ve{r})$ which are centered around the $L$ lattice sites~$i$ and
carry the spin-orbital index $\sigma$. 
With this basis, the 
Hamiltonian~(\ref{2.290}) can be written as
\begin{equation}
\label{er1}
\hat{H}^{\rm eff}_0=\sum_{i,j} \sum_{\sigma,\sigma'}
t^{\sigma,\sigma'}_{i,j}\!\!\left\{n(\ve{r})\right\}\hcd_{i,\sigma}\hc_{j,\sigma'}\;.
\end{equation}
Here, the `electron transfer' or `hopping' parameters 
\begin{equation}
\label{tzr}
t^{\sigma,\sigma'}_{i,j}\!\!\left\{n(\ve{r})\right\}
\equiv 
\int {\rm d}^3r
\phi^*_{i,\sigma}(\ve{r})\Bigg(
-\frac{\Delta_{\ve{r}}}{2m}+V(\ve{r})+
e^2\int {\rm d}^3r'\frac{n(\ve{r}')}{|\ve{r}-\ve{r}'|}
+V^{\rm KS}_{\rm xc}\!\left\{n(\ve{r})\right\}\Bigg)
\phi_{j,\sigma}(\ve{r})
\end{equation}
depend on the particle density
 \begin{equation}\label{pden}
n(\ve{r})=\sum_{i,j} \sum_{\sigma,\sigma'}\phi^*_{i,\sigma}(\ve{r})
\phi_{j,\sigma'}(\ve{r})\langle
 \hcd_{i,\sigma}\hc_{j,\sigma'}
 \rangle_{\Psi_0} \; ,
\end{equation}
where $|\Psi_0\rangle$ is the ground state of~(\ref{er1}),  
\begin{equation}\label{er4}
\hat{H}^{\rm eff}_0|\Psi_0\rangle=
E_0 |\Psi_0\rangle\;.
\end{equation}
At least in principle, the self-consistent solution of 
the `Kohn--Sham equations' (\ref{er1})-(\ref{er4}) is the central 
part of most DFT applications in solid-state physics.
Note, however, that actual numerical implementations of the DFT 
usually do not work with Wannier functions but use atomic orbitals 
or plane waves as basis sets. 

Despite the rather drastic approximations which have led to the
Kohn--Sham equations,
a comparison of theoretical and experimental results
has revealed a remarkable agreement for a 
large number of materials. Therefore, the LDA has become the 
most important tool for the investigation of electronic properties in 
solid-state physics. 
There are, however, well-known problems with certain 
classes of materials. For example, band gaps in insulators or 
semiconductors are usually found to be significantly smaller in DFT 
than in experiment. 
Even bigger discrepancies arise for materials with strong local Coulomb 
interactions. These are, 
in particular, transition metals, lanthanides, and their respective 
compounds. Such systems have been investigated in the past
mostly based on model systems, which we discuss in the following section.

\subsection{Multi-Band Hubbard models} 
\label{sec3}

We distinguish `localised' orbitals, $\sigma \in \ell$,
and `delocalised' orbitals,   $\sigma \in {\rm d}$, where
the localised orbitals are those which require a more 
sophisticated treatment of the local Coulomb interaction  
than provided by the LDA. The natural starting point for such a treatment
 is a multi-band Hubbard model of the form
\begin{eqnarray}
\label{h2}
\hat{H}_{\rm H}&=&\hat{H}_{0}+\sum_i\hat{H}_{i;{\rm c}}\;,\\\label{h2bb}
\hat{H}_{0}&\equiv& \sum_{i\neq j} \sum_{\sigma,\sigma'}
t^{\sigma,\sigma'}_{i,j} \hcd_{i,\sigma}\hc_{j,\sigma'}
+\sum_i\sum_{\sigma,\sigma' \in {\rm d}} \epsilon_i^{\sigma,\sigma'}
\hcd_{i,\sigma}\hc_{i,\sigma'} \;, \\[3pt]
\label{er}
\hat{H}_{i;{\rm c}}&\equiv&
\sum_{\sigma,\sigma' \in \ell} \epsilon_i^{\sigma,\sigma'}
\hcd_{i,\sigma}\hc_{i,\sigma'}
+\sum_{\sigma_1,\sigma_2,\sigma_3,\sigma_4 \in \ell}
U_{i}^{\sigma_1,\sigma_2,\sigma_3,\sigma_4}
\hcd_{i,\sigma_1} \hcd_{i,\sigma_2}\hc_{i,\sigma_3} 
\hc_{i,\sigma_4}\;.
\end{eqnarray}
This model contains 
a general two-particle interaction in the localised
orbitals, and fixed hopping parameters, $t^{\sigma,\sigma'}_{i,j}$, 
and orbital energies, $\epsilon_{i}^{\sigma,\sigma'}=t^{\sigma,\sigma'}_{i,i} $.
Since the parameters  $t^{\sigma,\sigma'}_{i,j}$ are usually derived from a 
DFT calculation, see Eq.~(\ref{tzr}), they
already contain the   Coulomb interaction on a DFT level. 
For the localised orbitals this means that, through the on-site energies
$\epsilon_{i}^{\sigma,\sigma'}$, the Coulomb interaction appears 
twice in the Hamiltonian $\hat{H}_{i;{\rm c}}$. We will address this 
 so-called 
 `double-counting problem' in Sec.~\ref{gdft}. 
  
 In the context of the Gutzwiller variational theory we
 need the eigenstates $\ket{\Gamma}_i$ and
 the eigenvalues $E_{i,\Gamma}$ 
of the Hamiltonian $\hat{H}_{i;{\rm c}}$. They allow us to write 
 $\hat{H}_{i;{\rm c}}$ as 
\begin{equation}\label{rtzu}
\hat{H}_{i;{\rm c}}=\sum_{\Gamma} E_{i,\Gamma} \hat{m}_{i,\Gamma}
\;\;,\;\;
\hat{m}_{i,\Gamma}\equiv\ket{\Gamma}_{i\;i}\!\bra{\Gamma}\;.
\end{equation}
As a simple example, we consider a model with  two degenerate
 $e_{\rm g}$ orbitals in a cubic environment. In this case, we may set 
$\epsilon^{\sigma,\sigma'}_i=0$ and the local 
Hamiltonian then has the form
\begin{eqnarray}\label{2.160a}
\hat{H}_{i;{\rm c}} &=&
U \sum_{e}\hat{n}_{e,\uparrow}\hat{n}_{e,\downarrow}
+U'\sum_{s,s'}\hat{n}_{1,s}\hat{n}_{2,s'}
-J\sum_{s}\hat{n}_{1,s}\hat{n}_{2,s}
\label{twoorbhamiltonian} \\[3pt]
&& +J\sum_{s}\hat{c}_{1,s}^{\dagger}
\hat{c}_{2,\bar{s}}^{\dagger}
\hat{c}_{1,\bar{s}}^{\vphantom{+}}
\hat{c}_{2,s}^{\vphantom{+}}
+J_C \Bigl(
\hat{c}_{1,\uparrow}^{\dagger}\hat{c}_{1,\downarrow}^{\dagger}
\hat{c}_{2,\downarrow}^{\vphantom{+}}\hat{c}_{2,\uparrow}^{\vphantom{+}}
+ 
\hat{c}_{2,\uparrow}^{\dagger}\hat{c}_{2,\downarrow}^{\dagger}
\hat{c}_{1,\downarrow}^{\vphantom{+}}\hat{c}_{1,\uparrow}^{\vphantom{+}}
\Bigr)\;,  \nonumber
\end{eqnarray}
where $e=1,2$ labels the $e_{\rm g}$ orbitals, $s=\uparrow,\downarrow$ 
is the spin index and we use the convention $\bar{\uparrow}\equiv \downarrow$, 
$\bar{\downarrow}\equiv\uparrow$.
For $e_{\rm g}$ orbitals,
only two of the three parameters in~(\ref{2.160a}) are independent 
 since the symmetry relations $U'=U-2J$ and $J=J_C$ hold.
 In our model, we have four spin-orbital states $\sigma=(e,s)$  per atom, 
leading to a $2^4=16$-dimensional atomic Hilbert space. 
All eigenstates~$|\Gamma\rangle_i $ of $\hat{H}_{i;{\rm c}}$ 
with particle numbers $N\neq 2$ 
 are simple Slater determinants of spin-orbital states~$|\sigma\rangle$
  and their energies are
\begin{equation}
\begin{array}{ll}
E_{\Gamma}=0 & (N=0,1)\;,\\
E_{\Gamma}=U+2U'-J & (N=3)\;,\\
E_{\Gamma}=2U+4U'-2J & (N=4)\;.
\end{array}
\end{equation}

\begin{table}[t]
\centering
\begin{tabular}{c|c|c|c}
\# & Atomic eigenstate $|\Gamma \rangle $ & Symmetry  & energy $E_{\Gamma} $ \\
\hline 
1 & $|\uparrow ,\uparrow \rangle $ & ${}^3A_2$  & $U'-J$  \\ 
2 & $(|\uparrow ,\downarrow \rangle +|\downarrow ,\uparrow \rangle )/\sqrt{2}
$ & ${}^3A_2$  & $U'-J$  \\ 
3 & $|\downarrow ,\downarrow \rangle $ & ${}^3A_2$ 
& $U'-J$  \\ 
4 & $(|\uparrow ,\downarrow\rangle -|\downarrow ,\uparrow \rangle )/\sqrt{2}
$ & ${}^1E$ &  $U'+J$  \\ 
5 & $(|\uparrow \downarrow ,0\rangle -|0,\uparrow \downarrow \rangle )/\sqrt{2}$
 & ${}^1E$ &  $U-J_{\text{C}}$\\ 
6 & $(|\uparrow \downarrow ,0\rangle +|0,\uparrow \downarrow \rangle )/\sqrt{2}$
 & ${}^1A_1$  & $U+J_{\text{C}}$  \\ 
\end{tabular}
\caption{Two-particle eigenstates with symmetry specifications and energies.
\label{tableone}}
\end{table}

The two-particle eigenstates are slightly more 
 complicated because some of them are 
 linear combinations of Slater determinants. 
We introduce the basis
 \begin{equation}\label{2.299a}
|s,s'\rangle\equiv\hcd_{1,s}\hcd_{2,s'}|0\rangle\;\;,\;\;
|\!\uparrow\downarrow,0\rangle\equiv\hcd_{1,\uparrow}
\hcd_{1,\downarrow}|0\rangle\;\;,\;\;
|0,\uparrow\downarrow\rangle\equiv\hcd_{2,\uparrow}
\hcd_{2,\downarrow}|0\rangle\;
\end{equation}
of two-particle states, which are used to set up 
 the eigenstates of $\hat{H}_{i;{\rm c}}$, see table~\ref{tableone}.
The states of lowest energy are the three triplet states 
with spin $S=1$, which belong to the 
 representation $A_2$ of the cubic point-symmetry group. Finding 
 a high-spin ground state is a simple consequence of 
Hund's first rule. Higher in energy 
 are the two degenerate singlet states of symmetry $E$ 
and the non-degenerate singlet state of symmetry $A_1$.  

The eigenstates of the local Hamiltonian $\hat{H}_{i;{\rm c}}$ play an 
essential role 
 in the formulation of the multi-band Gutzwiller theory in Sec.~\ref{gw}.
 Since in most applications only a finite (and not too large) number of 
 localised orbitals is taken 
 into account, these eigenstates can be readily calculated by 
standard numerical  techniques. The special case of a 
$3d$-shell in a cubic environment has been discussed analytically 
in great detail in the textbook by Sugano, Ref.~\cite{sugano1970}.   

\section{Gutzwiller wave functions}\label{gw}
{\bf \large The single-band Hubbard model}
\vspace{0.2cm}

To understand the main physical idea behind the Gutzwiller variational theory 
it is instructive to start with a consideration of the 
 single-band Hubbard model 
 \begin{equation}\label{h3}
\hat{H}_{1{\rm B}}=\sum_{i,j} \sum_{s=\uparrow,\downarrow}
t_{i,j} \hcd_{i,s}\hc_{j,s}+
U\sum_{i}\hat{d}_i\;\;, 
 \;  \;\hat{d}_i\equiv \hat{n}_{i,\uparrow}\hat{n}_{i,\downarrow}\;.
\end{equation} 
In Hartree--Fock theory, 
 one uses a  variational wave function which is a one-particle product state 
\begin{equation}\label{4.10}
\ket{\Psi_0}=\prod_{\gamma}\hhd_{\gamma}\ket{0}\;.
\end{equation} 
in order to investigate many-particle Hamiltonians such as~(\ref{h3}). 
It is well known, however, that such wave functions 
are insufficient 
for systems with 
 medium to strong Coulomb interaction effects, see, e.g, our later discussion 
in  Sec.~\ref{appl}.
 It is a particular problem of a Hartree--Fock treatment that local 
 charge fluctuations can only be suppressed in that approach 
 by a spurious breaking of symmetries. Therefore, it usually overestimates
the stability 
 of phases with a broken symmetry.
Hartree--Fock wave functions, however, 
 can still be a reasonable starting point in order to set up more
 sophisticated variational wave functions. This leads us to the general class of 
 `Jastrow wave functions'  
\cite{feenberg1969,clements1993}, which are defined as
\begin{equation}\label{4.20}
 \ket{\Psi_{\rm J}}=\hat{P}_{\rm J} \ket{\Psi_0}\;.
\end{equation}
Here, $\ket{\Psi_0}$ is again a one-particle product state 
and $\hat{P}_{\rm J} $ is a correlation operator, which can be chosen 
in various ways in order to minimise the variational ground-state energy.
The `Gutz\-willer wave function' (GWF) is a special 
 Jastrow wave function with a particular choice of the correlation 
 operator $\hat{P}_{\rm J}$. It was introduced by 
Gutzwiller~\cite{gutzwiller1963,gutzwiller1964,gutzwiller1965} 
in the form
\begin{equation}\label{4.30}
\ket{\Psi'_{\rm G}}\equiv\hat{P}'_{\rm G}\ket{\Psi_0}=
\prod_{i}\hat{P}'_{i}\ket{\Psi_0}
\end{equation}
and with the purpose to study ferromagnetism in a single-band Hubbard model.
The (local) `Gutzwiller correlation operator'
\begin{equation}\label{4.35}
\hat{P}'_{i}\equiv g^{\hat{d}_i}=1-(1-g)\hat{d}_i\;,
\end{equation}
for each lattice site $i$ contains a variational parameter $g$  
 (with $0\leq g\leq 1$),  
which allows one to supress local double occupancies that are
 energetically unfavourable for a finite Hubbard interaction $U>0$. 

The Hilbert space of the local Hamiltonian for the one-band Hubbard model 
 is four-dimensional where a local basis $\ket{I}$ is given by the states
$\ket{\emptyset}$, $\ket{\uparrow}$, $\ket{\downarrow}$, and $\ket{d}$
 for  empty, singly-occupied and doubly-occupied  sites, respectively. 
By working with the occupation operator 
$\hat{d}_i$ in~(\ref{4.35}),
 Gutzwiller  singled out the doubly-occupied state $\ket{d}$. A more 
 symmetric definition of the {\sl local}
 Gutzwiller correlator~(\ref{4.35}) is given by
\begin{equation}\label{4.50}
\hat{P}_i=\prod_{I}\lambda^{\hat{m}_{i,I}}_{I}=
\sum_{I}\lambda_{I}\hat{m}_{i,I}
\end{equation}
where the operators $\hat{m}_{i,I}=\ket{I}_{i\;i}\!\bra{I}$ are the projectors 
 onto the four atomic eigenstates $\ket{I}$. The operator~(\ref{4.50})
 contains four parameters $\lambda_{I}$ instead of only one parameter 
 $g$ in Gutzwiller's definition~(\ref{4.35}). It can be readily shown, 
however, that the operators~(\ref{4.35}) 
and~(\ref{4.50}) define the same sets of variational 
 wave functions as long as the respective one-particle states 
$\ket{\Psi_0}$ are also treated as variational objects. Therefore, the 
  wave functions, defined by~(\ref{4.50}),
 contain more variational parameters than are actually needed. This 
 surplus of parameters will turn out to be quite useful when we evaluate 
 expectation values in the limit of infinite spatial dimensions.
 Moreover, for the multi-band generalisation of Gutzwiller wave functions
   in the following section, Eq.~(\ref{4.50}) is the most  
natural starting point. 

\vspace{0.3cm}
{\bf \large Multi-band Hubbard models}
\vspace{0.2cm}

 It is pretty obvious \cite{buenemann1998,buenemann2005} how  the Gutzwiller 
wave functions~(\ref{4.30}) can be generalised for the 
 investigation of the multi-band Hubbard models~(\ref{h2}). 
The starting 
 point is again a (normalised) single-particle product state $\ket{\Psi_0}$
 to which we apply a Jastrow factor that is a product 
 of local correlation operators. Hence, the multi-band Gutzwiller
 wave functions are given as 
 \begin{equation}\label{4.200}
\ket{\Psi_{\rm G}}=\hat{P}_{\rm G}\ket{\Psi_0}=
\prod_i\hat{P}_{i}\ket{\Psi_0}\;,
\end{equation}
where, as in~(\ref{4.50}), we might work with a local correlation 
 operator of the form
 \begin{equation}\label{4.201}
\hat{P}_{i}=\sum_{\Gamma_i}\lambda_{i;\Gamma_i}
\hat{m}_{i;\Gamma_i}\;\;,\;\;\hat{m}_{i;\Gamma_i}
=\ket{\Gamma}_{i\;i}\!\bra{\Gamma}\;.
 \end{equation}
The variational parameters $\lambda_{i;\Gamma_i}$
allow us to optimise the occupation of each 
eigenstate $\ket{\Gamma}$ of the local Hamiltonian $\hat{H}_{i;{\rm c}}$. 
 In multi-orbital systems, however, these  states are
 usually degenerate and not uniquely defined. Moreover,
  it is not clear whether, in a solid, the (atomic) 
 eigenstates $\ket{\Gamma}_i$ lead to the best variational ground state 
 of the form~(\ref{4.200}). Instead of~(\ref{4.201}) it may therefore be 
 better to work with the general local correlation operator
 \begin{equation}\label{4.210}
\hat{P}_{i}=\sum_{\Gamma_i,\Gamma'_i}\lambda_{i;\Gamma_i,\Gamma_i'}
\ket{\Gamma}_{i\;i}\!\bra{\Gamma'}
\;,
 \end{equation}
which contains a matrix $\lambda_{i;\Gamma_i,\Gamma_i'}$  of variational 
 parameters. The analytical evaluation of expectation values, which 
 we discuss in the following section, can be carried out without 
 additional efforts for the general correlation operator~(\ref{4.210}).
 In numerical applications, however, we often have to restrict 
 ourself to the simpler operator~(\ref{4.201}) since the 
  number of parameters $\lambda_{i;\Gamma_i,\Gamma_i'}$ may become 
 prohibitively large. Alternatively, one can try to identify the  
 `relevant' non-diagonal elements of $\lambda_{i;\Gamma_i,\Gamma_i'}$ and
 take only these into account.
 Such strategies have been discussed in more detail in 
Ref.~\cite{buenemann2012b}. 

 For systems without superconductivity, the Gutzwiller wave function 
 should be an eigenstate of the total particle number operator 
\begin{equation}
\hat{N}=\sum_{i,\sigma}\hat{n}_{i,\sigma}\;.
\end{equation}
 This requires that $\hat{N}$ commutes with $\hat{P}_{\rm G}$, which leads to
\begin{equation}\label{4.240}
\sum_{\Gamma,\Gamma'}\lambda_{i;\Gamma,\Gamma'}(|\Gamma|-|\Gamma'|)
\ket{\Gamma}_{i\;i}\!\bra{\Gamma'}=0
\end{equation}
 where $|\Gamma|$ is the number of particles in the state $\ket{\Gamma}_i$.
  From equation~(\ref{4.240}), we conclude that $\lambda_{i;\Gamma,\Gamma'}$
 can only be finite for states $\ket{\Gamma}_i,\ket{\Gamma'}_i$ with the 
 same particle number. 
In a similar way, one can show that
  these states have to belong to the same representation of the 
 point symmetry group.
 To study superconducting systems, one works with BCS-type one-particle 
 wave functions $\ket{\Psi_0}$ for which the particle number is not conserved.
 In this case, the variational-parameter matrix $\lambda_{i;\Gamma,\Gamma'}$ 
 has to be finite also for states $\ket{\Gamma}_i,\ket{\Gamma'}_i$ 
with different  particle numbers, see Refs.~\cite{buenemann2005,buenemann2005b}. 

To keep notations simple in this tutorial presentation, 
we will restrict ourself 
 to the case of a diagonal and real variational-parameter matrix and do not
 consider superconducting states. Consequently, the local correlation 
 operators are Hermitian, $\hat{P}_i^{\dagger}=\hat{P}_i$.  
Moreover, we work with a 
 spin-orbital basis $\sigma$ for which the non-interacting local 
 density matrix
\begin{equation}\label{8.12a}
C_{i;\sigma,\sigma'}\equiv\langle \hcd_{i,\sigma}\hc_{i,\sigma'}\rangle_{\Psi_0}
\end{equation}
 is diagonal,
\begin{equation}\label{8.12}
C_{i;\sigma,\sigma'}
=\delta_{\sigma,\sigma'}n^0_{i,\sigma}\;.
\end{equation}
This can always  be achieved (i.e., for any $\ket{\Psi_0}$) by a proper
 transformation of the local basis~$\sigma$. To simplify the notations 
further, we will frequently drop lattice-site indices in purely local 
 equations.  
\section{Gutzwiller energy functional in infinite dimensions}\label{enfunc}
The evaluation of expectation values for Gutzwiller wave functions 
remains a difficult many-particle problem even in the simplest case 
of a single-band Hubbard model.
 It has been achieved for this model in one dimension both for paramagnetic 
  and for ferromagnetic states~\cite{metzner1988,kollar2002,gebhard1987,gebhard1988,gebhard1990}. In the opposite 
 limit of infinite spatial dimensions,  expectation values can be evaluated 
 for the general class of wave-functions~(\ref{4.200}). In this section, 
 we summarise the main technical ideas behind this evaluation and 
 discuss the resulting energy functional. An application of this 
 functional to finite-dimensional systems is usually denoted as 
 the `Gutzwiller approximation'  because, for the single-band model, Gutzwiller 
 has derived the very same functional\cite{gutzwiller1963,gutzwiller1964,gutzwiller1965} by means of  combinatorial 
techniques \cite{buenemann1998b}.     
\subsection{Diagrammatic expansion}
In order to determine the expectation value 
\begin{equation}\label{8.1222}
\langle \hat{H}_{\rm H} \rangle_{\Psi_{\rm G}}=
\frac{\langle\Psi_{\rm G}|  \hat{H}_{\rm H}|\Psi_{\rm G} \rangle   }
{\langle\Psi_{\rm G}| \Psi_{\rm G} \rangle  }
\end{equation}
of the Hamiltonian~(\ref{h2})
we need to evaluate the following quantities ($i\ne j$)
\begin{eqnarray}\label{sdf1}
\langle\Psi_{\rm G}| \hcd_{i,\sigma}\hc_{j,\sigma'} |\Psi_{\rm G} \rangle &=&
\Bigl \langle (\hat{P}_{i}  \hcd_{i,\sigma} \hat{P}_{i})
(\hat{P}_{j}  \hc_{j,\sigma'} \hat{P}_{j})
\prod_{l\neq (i,j))}\hat{P}^2_{l}\Bigr \rangle_{\Psi_0}\;,\\\label{sdf2}
\langle\Psi_{\rm G}| \hat{m}_{i;\Gamma} |\Psi_{\rm G} \rangle &=&
\Bigl \langle (\hat{P}_{i} \hat{m}_{i;\Gamma} \hat{P}_{i})
\prod_{l\neq i}\hat{P}^2_{l}\Bigr \rangle_{\Psi_0}\;,\\\label{sdf3}
\langle\Psi_{\rm G}|\Psi_{\rm G} \rangle &=&
\Bigl \langle \prod_{l}\hat{P}^2_{l}\Bigr \rangle_{\Psi_0}\;.
\end{eqnarray}
The r.h.s.\ of all three equations~(\ref{sdf1})-(\ref{sdf3})
 can be evaluated by means of Wick's theorem
 because the wave function $\ket{\Psi_0}$ is a single-particle product state.
 In this way, we can represent all contributions by diagrams
 with `internal vertices' $l$ (from operators $\hat{P}^2_{l}$),
  `external vertices' $i$ in Eq.~(\ref{sdf2}) (or $i$ and $j$ in 
Eq.~(\ref{sdf1})) and lines 
\begin{equation}\label{pl}
P^{\sigma,\sigma'}_{l,l'}\equiv 
\langle\hat{c}^{\dagger}_{l,\sigma}\hat{c}_{l',\sigma'}^{\phantom{\dagger}}\rangle_{\Psi_0}
\end{equation}
which connect these vertices. This diagrammatic expansion, however,
 is still very complicated even in the limit of infinite spatial dimensions.
 As shown in more detail in Refs.\cite{buenemann1998,buenemann2005}, it is 
  very beneficial 
in this limit 
to introduce the (local) constraints
\begin{eqnarray} \label{8.140a}   
  1&=& \langle \hat{P}^{2}_l\rangle_{\Psi_0} \; ,  
\\  \label{8.140b}   
\langle \hcd_{l,\sigma} \hc_{l,\sigma'}   
\rangle_{\Psi_0}  &=&
\langle \hcd_{l,\sigma}  \hat{P}^{2}_l  
\hc_{l,\sigma'} \rangle_{\Psi_0}   \;.
\end{eqnarray}
These constraints do not restrict our total set of 
 variational wave functions~(\ref{4.200})  because they merely exploit 
 the fact that we have introduced more variational parameters 
$\lambda_{\Gamma}$ than are actually needed; see the discussion on the 
single-band model in Sec.~\ref{gw}. Note that moving the 
operator $\hat{P}^{2}_l$ relative to $\hcd_{l,\sigma}$ and  
$\hc_{l,\sigma'} $ does not alter 
 the whole set of constraints~(\ref{8.140a}),~(\ref{8.140b}). 

\begin{SCfigure}[1]\label{fig1}
 \centering
 \includegraphics[width=0.55\textwidth]{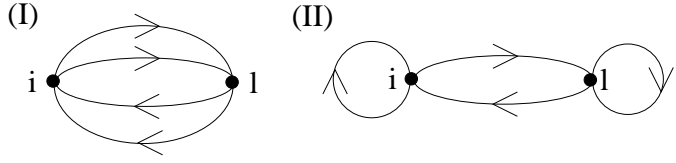}
 \caption{Two examples of double-occupancy diagrams ($l \ne i$). Diagram (II) vanishes due
 to the constraints~(\ref{8.140a}),~(\ref{8.140b}). Diagram (I) vanishes in the
 limit of infinite spatial dimensions.}
\end{SCfigure}

The constraints~(\ref{8.140a}),~(\ref{8.140b}) have an important 
 consequence: Each diagram that results 
from~(\ref{sdf1})-(\ref{sdf3}) is non-zero only when 
 all of its internal vertices $l$ are connected to other
 vertices by at least four lines. As a simple example, 
Fig.~\ref{fig1} shows two (first order) 
 diagrams which result from~(\ref{sdf2}). While the 
 constraints do not affect diagram (I), they ensure that diagram (II) 
vanishes.   

In order to have a meaningful (i.e., finite) kinetic energy per 
 lattice site our lines have to vanish like 
 \begin{equation}\label{7.130}
P^{\sigma,\sigma'}_{i,j} \sim 
\frac{1}{\sqrt{2D}^{||i-j||}}\;,
\end{equation}
on a hyper-cubic lattice in the limit of large spatial dimensions $D$. 
 Here, we introduced the `New-York metric'
\begin{equation}\label{7.40}
||i-j||\equiv \sum^{D}_{k=1}|R_{i;k}-R_{j;k}| \;,
\end{equation}
where $R_{i;k}$ is the $k$-th component of the lattice site vector
 $\ve{R}_{i}$. Note that the number of neighbouring sites with distance 
$||i-j||$ 
 is given by
\begin{equation}\label{7.40vv}
N^{||i-j||}_{\rm n.n.}=2D^{||i-j||} \;.
\end{equation}
The scaling behaviour~(\ref{7.130}) in infinite dimensions
 has significant consequences for our diagrammatic expansion. As an 
 example, we consider diagram (I) in Fig.~\ref{fig1}. If we skip, for simplicity,
 any spin-orbital dependence of lines, this diagram leads to the contribution  
 \begin{equation}
{\rm diagram\;(I)}\sim\sum_lP_{i,l}^4=\mathcal{O}\left(\frac{1}{D}  \right)
\stackrel{D\to \infty}{\longrightarrow}0\;.
 \end{equation}
where the scaling $1/D$ results from equations~(\ref{7.130}) and~(\ref{7.40vv}).
 In general, one can show that in infinite dimensions a diagram vanishes
 if it contains an internal vertex that is connected  to other vertices by three or more lines. The constraints~(\ref{8.140a}),~(\ref{8.140b}) ensure 
that this is the case 
 for {\sl all} diagrams which contain at least one internal vertex.    
 Our arguments,  so far, only apply to diagrams in which all internal vertices
  are connected to the external vertices $i$ (or $i$ and $j$).
  Of course, if we apply Wick's theorem to equations~(\ref{sdf1})-(\ref{sdf2})
 we also obtain diagrams with internal vertices that are 
 not connected to the external vertices. These diagrams, however,
 are exactly cancelled by the norm diagrams from Eq.~(\ref{sdf3})
 as shown in Ref.~\cite{buenemann1998}. In summary, we therefore end up 
 with the simple results 
 \begin{eqnarray}\label{sdf1b}
\langle \hcd_{i,\sigma}\hc_{j,\sigma'} \rangle_{\Psi_{\rm G}} &=&
\Bigl \langle (\hat{P}_{i}  \hcd_{i,\sigma} \hat{P}_{i})
(\hat{P}_{j}  \hc_{j,\sigma'} \hat{P}_{j})\Bigr \rangle_{\Psi_0}\;,\\\label{sdf2b}
\langle \hat{m}_{i;\Gamma} \rangle_{\Psi_{\rm G}} &=&
\Bigl \langle (\hat{P}_{i} \hat{m}_{i;\Gamma} \hat{P}_{i})\Bigr \rangle_{\Psi_0}\;,
\end{eqnarray} 
 for expectation values in the limit of infinite 
spatial dimensions.
These expectation values and the constraints~(\ref{8.140a}),~(\ref{8.140b}) will
 be further analysed in the following section and 
  lead us to the Gutzwiller energy functional in infinite dimensions. 

\subsection{Energy functional for multi-band systems}\label{enfunc2}

{\bf Notations}
\vspace{0.1cm}

We assume that the $2N$ (localised) spin-orbital
 states $\sigma$ are  ordered in some arbitrary 
 way, $\sigma= 1,\ldots,2N$ where $N$ is the number of localised orbitals
 per lattice site. 
In order to set up a proper basis of the local Hilbert space which belongs to
 $\hat{H}_{i;{\rm c}}$, 
 we introduce the following notations for the $2^{2N}$ possible 
electronic configurations (`Slater determinants').
\begin{itemize}
\item[i)\,] An atomic configuration $I$ is characterised by the electron 
occupation of the orbitals,
\begin{equation}\label{4.20a}
 I \in  \{\emptyset;
(1),\ldots,(2N);
(1,2),\ldots,(2,3),\ldots (2N-1,2N);
\ldots;
(1,\ldots,2N)
\}\;,
\end{equation}
where the elements in each set $I=(\sigma_1,\sigma_2,\ldots)$ 
are ordered, i.e., it is $\sigma_1<\sigma_2<\ldots$.
In general, we interpret the indices $I$ as sets in the usual 
 mathematical sense.
For example, in the atomic configuration
$I\backslash I'$ 
only those orbitals in $I$ that are not in $I'$ 
are occupied. The `complement' $\bar{I}$ is defined as
\begin{equation}
\bar{I}\equiv (1,\ldots,2N)\backslash I\;.
\end{equation}
where $(1,\ldots,2N)$ is the state with the maximum number of $2N$ electrons.
\item[ii)\,] A state with a specific configuration $I$ is given as 
\begin{equation}\label{4.30a}
\ket{I}=\hat{C}_{I}^{\dagger}\ket{0}\equiv\prod_{\sigma \in I}\hcd_{\sigma}\ket{0}=
\hcd_{\sigma_1}\dots\hcd_{\sigma_{|I|}}\ket{0}\;,
\end{equation}
where the operators $\hcd_{\sigma}$ are in ascending order, i.e., it is 
 $\sigma_1<\sigma_2\ldots<\sigma_{|I|}$ and $|I|$ is the number of particles
 in $I$. Products of annihilation operators, such as
\begin{equation}\label{4.35a}
\hat{C}_{I}^{}\equiv\prod_{\sigma\in I}\hc_{\sigma}=\hc_{\sigma_1}\dots\hc_{\sigma_{|I|}},
\end{equation}
will be  placed in descending order, i.e., with 
$\sigma_1>\sigma_2\ldots>\sigma_{|I|}$. Note that we have introduced  the operators
 $\hat{C}_{I}^{\dagger}$ and  $\hat{C}_{I}^{}$ just as convenient abbreviations.
 They must not be misinterpreted as 
  fermionic creation or annihilation operators. The `sign function'
\begin{equation}
f(\sigma,I)\equiv \langle I\cup \sigma| \hcd_\sigma |I \rangle
\end{equation}
counts whether an odd or even number of commutations is required 
 to place $\sigma$ in its proper position in $I$ ($f(\sigma,I)=\mp 1$). 
It vanishes if $\sigma \in I$. 

\item[iv)\,] The operator $\hat{m}_{I,I'}\equiv \ket{I}\bra{I'}$ 
describes the transfer
 between configurations $I'$ and $I$. It can be written as  
 \begin{equation}\label{4.50a}
\hat{m}_{I,I'}=
\hat{C}_{I}^{\dagger}
\hat{C}_{I'}^{}
\prod_{\sigma''\in J}(1-\hat{n}_{\sigma''})
\end{equation}
where $J\equiv \overline{I\cup I'}$. A special case,
  which derives from~(\ref{4.50a}), is the occupation operator
\begin{equation}\label{4.52a}
\hat{m}_{I}\equiv  \ket{I}\bra{I}=\prod_{\sigma\in I}\hat{n}_{\sigma}
\prod_{\sigma'\in \bar{I}}(1-\hat{n}_{\sigma'})\;.
\end{equation}
\end{itemize}
The states $\ket{I}$ form a basis of the atomic Hamitonian's 
Hilbert space. Therefore,
 we can write the eigenstates of the local Hamiltonian~(\ref{rtzu}) as
\begin{equation}\label{4.60a}
|\Gamma\rangle =\sum_{I}T_{I,\Gamma}\ket{I}
\end{equation}
with coefficients $T_{I,\Gamma}$. For a simple example, see the two-particle 
 states in table~\ref{tableone}. 

\vspace{0.3cm}
{\bf Local energy}
\vspace{0.1cm}

The evaluation of the r.h.s.\ of Eq.~(\ref{sdf2b}) is straightforward 
if we use
 \begin{equation}
\hat{P} \hat{m}_{\Gamma} \hat{P}=\lambda^2_{\Gamma}\hat{m}_{\Gamma}\;.
\end{equation}
This equation gives us 
 \begin{equation}\label{eq52}
\langle \hat{m}_{\Gamma}  \rangle_{\Psi_{\rm G} }
=\lambda^2_{\Gamma}m^0_{\Gamma}\;,
\end{equation}
where 
 \begin{equation}
m^0_{\Gamma}=\langle \hat{m}_{\Gamma} \rangle_{\Psi_0}=\sum_{I}|T_{I,\Gamma}|^2m^0_{I}\;\;\;,\;\;\;
m^0_{I}=\prod_{\sigma \in I}n^0_{\sigma}\prod_{\sigma \notin I}(1-n^0_{\sigma})\;.
\end{equation}
Here we have used Eqs.~(\ref{4.201}),~(\ref{8.12}), and~(\ref{4.52a}).

\vspace{0.3cm}
{\bf Hopping expectation values}
\vspace{0.1cm}

For the evaluation of~(\ref{sdf1b}) we start with
 \begin{equation}\label{378}
\hat{P}  \hcd_{\sigma} \hat{P}=\sum_{\Gamma,\Gamma'}
\lambda_{\Gamma}\lambda_{\Gamma'}
\sum_{I_1,I'_1,I_2,I'_2} 
\langle I_2| \hcd_{\sigma}   |I_1' \rangle
T_{I_1,\Gamma}T^*_{I_2,\Gamma}T_{I'_1,\Gamma'}T^*_{I_2',\Gamma'}
\hat{m}_{I_1,I'_2}
 \end{equation}
which follows from Eqs.~(\ref{4.201}), (\ref{4.50a}), (\ref{4.60a}). 
 Note that the second operator $\hat{P}  \hc_{\sigma'} \hat{P}$ 
in~(\ref{sdf1b})  
is just the 
 conjugate of~(\ref{378}) with  $\sigma$ replaced by $\sigma'$.
Hence, the only remaining expectation values
 which we need to evaluate in~(\ref{sdf1b}) have the form
\begin{equation}\label{1po}
E(I,I';J,J')\equiv\langle  
\hat{m}_{i;I,I'}\hat{m}_{j;J,J'}
 \rangle_{\Psi_0}\;.
\end{equation}
An application of Wick's Theorem to~(\ref{1po}) leads, in general,
 to a number of diagrams with (potentially) several lines connecting
 the lattice sites $i$ and $j$. At this point, however, we again apply the
  infinite-dimensional rule that all diagrams with three or more lines 
 connecting $i$ and $j$ can be discarded. Hence, the only remaining 
 diagrams are those with exactly one line between $i$ and $j$. 
Together with Eq.~(\ref{8.12}),  
we therefore find
\begin{equation}
E(I,I';J,J')=\sum_{\gamma}f(\gamma,I')
\delta_{I,I'\cup\gamma}
\frac{m^0_{I'}}{1-n^0_{\gamma}}
\sum_{\gamma'}f(\gamma',J)
\delta_{J\cup\gamma',J'}
\frac{m^0_{J}}{1-n^0_{\gamma'}}
\langle \hcd_{i,\gamma}  \hc_{j,\gamma'} \rangle_{\Psi_0}\;.
\end{equation}
Altogether, we obtain the following result for the 
 hopping expectation value~(\ref{sdf1b}) in infinite dimensions
 \begin{equation}
\langle \hcd_{i,\sigma}\hc_{j,\sigma'} \rangle_{\Psi_{\rm G} }
=\sum_{\gamma,\gamma'}q_{\sigma}^{\gamma}q_{\sigma'}^{\gamma'}
\langle \hcd_{i,\gamma}  \hc_{j,\gamma'} \rangle_{\Psi_0}
\end{equation}
with the `renormalisation matrix' 
 \begin{eqnarray}\label{qf1}
q_{\sigma}^{\gamma}&=&\frac{1}{1-n^{0}_{\gamma}}
\sum_{\Gamma,\Gamma'}\lambda_{\Gamma}\lambda_{\Gamma'}
\sum_{I,I'}f(\sigma,I)f(\gamma,I')
T^*_{I\cup \sigma,\Gamma}T_{I,\Gamma'}
T^*_{I',\Gamma'}T_{I'\cup \gamma,\Gamma}m^0_{I'}\\\label{qf2}
&=&\frac{1}{n^{0}_{\gamma}}
\sum_{\Gamma,\Gamma'}\lambda_{\Gamma}
\lambda^{}_{\Gamma'}
\langle \Gamma|
\hcd_{\sigma}
|\Gamma'\rangle
 \Big  \langle
\big (|\Gamma  \rangle
\langle \Gamma' |  \hc_{\gamma}\big )
\Big  \rangle_{\Psi_0}\;.
\end{eqnarray}
\vspace{0.3cm}

{\bf  Constraints}
\vspace{0.1cm}

The explicit form of the 
 correlation operator~(\ref{4.201}), together with  Eq.~(\ref{8.12}), 
 gives us directly the explicit form of the 
constraints~(\ref{8.140a}),~(\ref{8.140b})
\begin{eqnarray}\label{5.5}
1&=&\sum_{\Gamma}
\lambda_{\Gamma}^{2}
\sum_{I}T_{I,\Gamma}T^*_{I,\Gamma}m^0_I\;,\\\label{5.5b}
\delta_{\sigma,\sigma'}n^0_{\sigma}&=&\sum_{\Gamma}
\lambda_{\Gamma}^{2}
\sum_{I(\sigma,\sigma'\in I)}f(\sigma,I\backslash \sigma)
f(\sigma',I\backslash \sigma')
T_{I\backslash \sigma,\Gamma}
T^*_{I\backslash \sigma',\Gamma}m^0_I\;.
\end{eqnarray}

\vspace{0.3cm}
{\bf Summary: Structure of the energy functional}
\vspace{0.1cm}

In summary, we obtain the following Gutzwiller energy functional for the 
 multi-band Hubbard models~(\ref{h2}) in infinite dimensions
\begin{eqnarray}
E^{\rm GA} &=& 
\sum_{i \ne j} 
\sum_{ \sigma,\sigma',\gamma,\gamma}
t^{\gamma,\gamma'}_{i,j} q_{\gamma}^{\sigma} 
\left( q_{\gamma'}^{\sigma'}\right)^{*} 
\langle \hat{c}_{i,\sigma}^{\dagger}\hat{c}_{j,\sigma'}^{\phantom{+}}\rangle_{\Psi_0}+
\sum_i\sum_{\sigma,\sigma' \in {\rm d}} \epsilon_i^{\sigma,\sigma'}
\langle \hcd_{i,\sigma}\hc_{i,\sigma'}\rangle_{\Psi_0}
\nonumber \\[3pt]
&&+ L\sum_{\Gamma}              
E_{\Gamma}\lambda^{2}_{\Gamma} m^0_{\Gamma} \;
\label{1.4b}
\end{eqnarray} 
where, for the delocalised orbitals, the renormalisation factors
 are $q_{\gamma}^{\sigma}=\delta_{\sigma,\gamma}$.  
  The single-particle state $|\Psi_0\rangle$ enters~(\ref{1.4b}) 
solely through the non-interacting density matrix $\tilde{\rho}$
with the elements 
\begin{equation}
\rho_{(i\sigma),(j\sigma')}\equiv
\langle \hat{c}_{j,\sigma'}^{\dagger}
\hat{c}_{i,\sigma}^{\phantom{+}}\rangle_{\Psi_0}\;.
\end{equation} 
Hence, the Gutzwiller 
energy functional simplifies to
\begin{equation} 
\label{1.4c}
E^{\rm GA}\left(\tilde{\rho},\lambda_{\Gamma}\right) =
\sum_{i \ne j}
\sum_{ \sigma,\sigma',\gamma,\gamma}
t^{\gamma,\gamma'}_{i,j} q_{\gamma}^{\sigma} 
\left( q_{\gamma'}^{\sigma'}\right)^{*} 
 \rho_{(j\sigma'),(i\sigma)}+
\sum_{i ; \sigma,\sigma' \in {\rm d}} \epsilon_i^{\sigma,\sigma'}
\rho_{(i\sigma),(i\sigma)} 
+ L\sum_{\Gamma}E_{\Gamma}\lambda^{2}_{\Gamma}m^0_{\Gamma} \;.
\end{equation} 
It has  to be minimised with respect to all elements of $\tilde{\rho}$
 and the variational parameters $\lambda_{\Gamma}$ obeying the 
 constraints~(\ref{5.5}),~(\ref{5.5b}) and 
\begin{equation}
\label{16} 
\tilde{\rho}^2=\tilde{\rho}\;.
\end{equation} 
The latter constraint ensures that $\tilde{\rho}$
belongs to a single-particle product state. 

There are several ways, how the constraints~(\ref{5.5}),~(\ref{5.5b})
 may be implemented in numerical calculations~\cite{buenemann2012b}. 
 In this tutorial introduction, we will simply assume 
 that Eqs.~(\ref{5.5}),~(\ref{5.5b}) are solved by
 expressing some of the parameters $\lambda_\Gamma$ by 
 the remaining `independent' parameters $\lambda^{\rm i}_\Gamma$. 
 Equation~(\ref{16}) is then the only remaining constraint 
in the minimisation of the  resulting energy function 
$\bar{E}^{\rm GA}\left(\tilde{\rho},\lambda^{\rm i}_{\Gamma}\right)$.
 If it is implemented 
 by means of Lagrange parameters, see  Appendix~\ref{appen1}, 
the minimisation with respect to
 $\tilde{\rho}$  leads to the effective single-particle Hamiltonian
\begin{equation}\label{5tg}
\hat{H}^{\rm eff}_0=\sum_{i\ne j}
\sum_{ \sigma,\sigma',\gamma,\gamma}
t^{\gamma,\gamma'}_{i,j} q_{\gamma}^{\sigma} 
\left( q_{\gamma'}^{\sigma'}\right)^{*} 
 \hcd_{i,\sigma}\hc_{j,\sigma'} 
+\sum_{i,\sigma\in d}\epsilon^{\sigma,\sigma'}_i\hcd_{i,\sigma}\hc_{i,\sigma}
+\sum_{i,\sigma\in {\rm \ell}}\eta_{\sigma}\hcd_{i,\sigma}\hc_{i,\sigma}\;
\end{equation}
which gives us $\ket{\Psi_0}$ as the ground state of~(\ref{5tg}),
\begin{equation}\label{5te}
\hat{H}^{\rm eff}_0|\Psi_0\rangle=E_0 |\Psi_0\rangle\;.
\end{equation}
The `fields' $\eta_{\sigma}$ for the localised orbitals 
in~(\ref{5tg}) are given as \cite{note2}
\begin{equation}\label{5th}
\eta_{\sigma}=\frac{\partial}{\partial n^0_{\sigma}}\bar{E}^{\rm GA}\left(\tilde{\rho},\lambda^{\rm i}_{\Gamma}\right)\;.
\end{equation}
The remaining numerical problem is the solution of 
Eqs.~(\ref{5tg})-(\ref{5th}) together with the minimisation condition
\begin{equation}
\frac{\partial}{\partial \lambda^{\rm i}_{\Gamma}}\bar{E}^{\rm GA}\left(\tilde{\rho},\lambda^{\rm i}_{\Gamma}\right)=0\;.
\end{equation}
Numerical strategies to solve these equations have been discussed 
 in detail in Ref.~\cite{buenemann2012b} to which we refer the interested 
reader.  

Up to this point, the effective single-particle Hamiltonian~(\ref{5tg}),   
\begin{equation}
\hat{H}^{\rm eff}_0=\sum_{\veck,\tau}E_{\veck,\tau}\hhd_{\veck,\tau}\hh_{\veck,\tau}\;,
\end{equation}
and its eigenvalues (`band-energies') 
$E_{\veck,\tau}$ in momentum space are just auxiliary 
 objects which determine $|\Psi_0\rangle$. One can readily show, however,
 that the non-interacting Fermi-surfaces, defined by the 
Fermi energy $E_{\rm F}$, 
\begin{equation}\label{eq:FS}
E_{\veck,\tau}-E_{\rm F}=0\;,
\end{equation}
are equal to the correlated Fermi surfaces because the 
momentum distribution
 \begin{equation} 
n_{\veck,\tau}\equiv 
\langle \hat{h}^{\dagger}_{\veck,\tau}\hat{h}^{}_{\veck,\tau} \rangle_{\Psi_{\rm G}}
\end{equation}
 has step discontinuities exactly at the momenta given by Eq.~(\ref{eq:FS}).
 The Fermi surface defined by~(\ref{eq:FS}) may therefore be compared 
 to those, e.g., from de-Haas-van-Alphen experiments. 
 Moreover, within a Landau-Fermi-liquid theory, 
 the eigenvalues $E_{\veck,\tau}$ turn out as the quasi-particle excitation
 energies which can be measured, e.g., in `angle-resolved photoemission
 spectroscopy' (ARPES) experiments, see 
Refs.~\cite{buenemann2005,buenemann2003b}.   

\subsection{Example: single-band Hubbard model} 
As a simple example we use the general results 
derived in Sec.~\ref{enfunc2}, to recover the well-known Gutzwiller energy 
functional for the 
 single-band Hubbard model~\cite{gutzwiller1963}. For this model, the 
(local) Gutzwiller
 correlation operator~(\ref{4.50}) had the form
  \begin{equation} 
\hat{P}=\lambda_{\emptyset}\hat{m}_{\emptyset}+
\lambda_{\uparrow}\hat{m}_{\uparrow}+\lambda_{\downarrow}\hat{m}_{\downarrow}
+\lambda_{d}\hat{d}\;,
\end{equation}
where
 \begin{eqnarray} 
\hat{m}_{\emptyset}&=&(1-\hat{n}_{\uparrow})(1-\hat{n}_{\downarrow})=
1-\hat{n}_{\uparrow}-\hat{n}_{\downarrow}+\hat{d}\;,\\
\hat{m}_{s}&=&\hat{n}_{s}(1-\hat{n}_{\bar{s}})=\hat{n}_{s}-\hat{d}\;,
\end{eqnarray}
$\bar{\uparrow}=\downarrow$,  $\bar{\downarrow}=\uparrow$, and $\hat{d}$
 has been defined in Eq.~(\ref{h3}).
 Equation~(\ref{eq52}) gives us the expectation value 
 for the occupation of the four local eigenstates,
\begin{eqnarray} \label{rt4a}
m_{\emptyset}&\equiv&\langle \hat{m}_{\emptyset}  \rangle_{\Psi_{\rm G}}
=\lambda^2_{\emptyset}(1-n^0_{\uparrow})(1-n^0_{\downarrow})\;,\\\label{rt4b}
\hat{m}_{s}&\equiv&\langle \hat{m}_{s}  \rangle_{\Psi_{\rm G}}=
\lambda^2_{s}n^0_{s}(1-n^0_{\bar{s}})\;,\\\label{rt4c}
d&\equiv&\langle \hat{d}  \rangle_{\Psi_{\rm G}}=
\lambda^2_{d}n^0_{\uparrow}n^0_{\downarrow}\;.
\end{eqnarray}
 With these equations, we can replace the original variational parameters 
 $\lambda_{\emptyset}$, $\lambda_{s}$, $\lambda_{d}$ by their corresponding 
 expectation values $m_{\emptyset}$, $m_{s}$, $d$. This simplifies the 
expressions for the constraints~(\ref{5.5}), (\ref{5.5b}) which then read  
\begin{eqnarray}\label{rt6}
1&=&m_{\emptyset}+m_{\uparrow}+m_{\downarrow}+d\;,\\\label{rt6b}
n^0_{s}&=&m_{s}+d\;.
\end{eqnarray}
Note that the second constraint~(\ref{rt6b}) simply ensures that 
 the correlated and the uncorrelated (spin-dependent) particle
 numbers are equal,
\begin{equation}
\langle \hat{n}_{s}  \rangle_{\Psi_{\rm G}}=m_{s}+d=n^0_{s}=
\langle \hat{n}_{s}  \rangle_{\Psi_{0}}
\;.
\end{equation}
Equations~(\ref{rt6}), (\ref{rt6b}) can be readily solved, e.g., 
 by expressing  $m_{\emptyset}$, $m_{s}$ as functions of $d$,
\begin{eqnarray}\label{rt6ii}
m_{\emptyset}&=&1-n^0_{\uparrow}-n^0_{\downarrow}+d\;,\\
m_{s}&=&n^0_{s}-d\;.
\end{eqnarray}
In this way, the only remaining variational parameter 
is the average number of doubly-occupied lattice sites $d$.

Finally, we can evaluate the hopping renormalisation factors~(\ref{qf2}),
\begin{eqnarray}\label{rt7}
q_s^{s'}(d)&=&\delta_{s,s'}
(\lambda_d\lambda_{\bar{s}}n^0_{\bar{s}}    +
\lambda_s \lambda_{\emptyset}(1-n^0_{\bar{s}}))\\
&=&\delta_{s,s'}\frac{1}{\sqrt{n^0_{s}(1-n^0_{s})}}
(\sqrt{m_{\bar{s}} d}+\sqrt{m_{s}m_{\emptyset}})\;,
\end{eqnarray}
 where, in the second line, we have used Eqs.~(\ref{rt4a})-(\ref{rt4c}).
 In summary, we obtain the variational energy functional
\begin{equation}\label{sbef}
\bar{E}^{\rm GA}(d,\Psi_0)=\sum_{s=\uparrow,\downarrow}(q^s_s(d))^2 \sum_{i,j}t_{i,j}
\langle \hcd_{i,s}  \hc_{j,s} \rangle_{\Psi_{0}}
+LUd
\end{equation}
for the single-band Hubbard model~(\ref{h3})  in the Gutzwiller approximation. 
\section{Applications}\label{appl}
\subsection{Ferromagnetism in a Two-Band Hubbard Model}\label{chap10.1} 
Since Gutzwiller's ground-breaking work we know that the 
single-band energy functional~(\ref{sbef})
  leads to ferromagnetic ground states 
only under very special circumstances, e.g., if the 
 density of states has a sharp peak at the Fermi level and the Coulomb 
 interaction is much larger than the band width.
From this observation, we can already conclude that ferromagnetism, as it 
 naturally appears in transition metals, is most probably related to the  
 orbital degeneracy of the partially filled 3d-shell in these systems.
 Therefore, it is quite instructive to study ferromagnetic instabilities 
 in a system with two orbitals, as a first step from the simple 
 one-band model towards a realistic description of materials with 
 partially filled  3d-shells.

\vspace{0.3cm}

{\bf A) Model Specification}
\vspace{0.1cm}

\begin{SCfigure}[1]\label{fig10.1}
{\centering
\includegraphics[width=9cm]{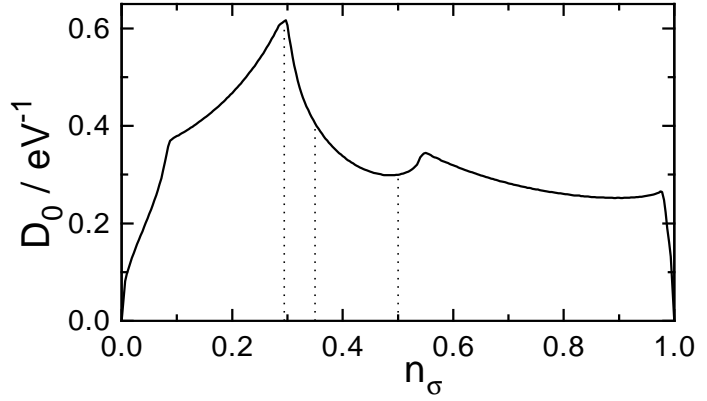}
\caption{Model density of states at the Fermi energy as a function of the
orbital filling $n_{\sigma}$. The dashed lines indicate the half-filled case 
($n_{\sigma}=0.5$  and the fillings
used in this section ($n_{\sigma}=0.29$ and  $n_{\sigma}=0.35$). 
The total bandwidth is $W=6.8$ eV.}}
\end{SCfigure}

 We consider a Hubbard model with two degenerate $e_{\rm g}$ orbitals 
per site on a simple three-dimensional cubic lattice. 
The local (atomic) Hamiltonian for this 
 system is given in equation~(\ref{2.160a}). 
 We include realistic hopping parameters 
for transition metal energy bands to the nearest and second-nearest neighbours
 in~(\ref{h2bb}).
 This
choice avoids pathological features in the energy bands, such as perfect
nesting at half band filling.
The single-particle part of the Hamiltonian~(\ref{h2bb}) 
is easily diagonalised in momentum space and leads to a density of states 
$D_0(\varepsilon)$ that is shown 
 as a function of the band filling in Fig.~\ref{fig10.1}.

The case $n_{\sigma}=n^0_{\sigma}= 0.29$ corresponds to a maximum in the density of 
states at the Fermi energy. For this band filling, we expect the strongest 
tendency towards ferromagnetism.

\vspace{0.3cm}

{\bf B) Variational Energy Functional}
\vspace{0.1cm}

For a two-band model, it is still possible to give a manageable explicit 
expression  of the energy as a function of the 
variational parameters.
 The eigenstates of the two-particle spectrum all belong to different 
 representations of the point symmetry group, see table 
\ref{tableone}.  Therefore, one can safely assume 
 that the variational-parameter matrix 
$\lambda_{\Gamma,\Gamma'}=\delta_{\Gamma,\Gamma'}$ is diagonal
 and we have  $\lambda_{\Gamma}=\lambda_{\Gamma'}$ for all 
 states $\ket{\Gamma}$, $\ket{\Gamma'}$ that are degenerate 
  due to the cubic symmetry. Then, we are left with 
$11$ (independent) variational parameters 
$m_{\Gamma}=\lambda^2_{\Gamma}m^0_{\Gamma}$: 
\begin{itemize}
\item[i)] two parameters for an empty and a fully occupied site: 
$m_{\emptyset}$, $f$;
\item[ii)] four parameters for 
singly and triply occupied sites: $m_{s}$ and $t_{s}$ with 
$s=\uparrow,\downarrow$;
\item[iii)] five parameters for 
doubly-occupied sites: $d^{\uparrow,\uparrow}_{t}$, $d^{\downarrow,\downarrow}_{t}$, $d^{0}_{t}$ (for the triplet ${}^3A_2$), $d_{E}$ (for the doublet ${}^1E$), 
and $d_{A}$ (for the singlet ${}^1A_1$).
  \end{itemize}

For our degenerate two-band model, the uncorrelated local 
density matrix~(\ref{8.12}) 
 is automatically diagonal and orbital independent,
\begin{equation}\label{11.20}
\big \langle \hcd_{i,(b,s)}  \hc_{i,(b',s')}\big \rangle_{\Psi_0}
=\delta_{s,s'} \delta_{b,b'}n^0_s\;.
\end{equation}  
As in the single-band case, the constraint 
equations~(\ref{5.5}),~(\ref{5.5b})
can be solved explicitly, e.g., by considering the occupations
\begin{eqnarray}\label{11.30a}
m_{\emptyset}&=&1-2n^0_{\uparrow }-2n^0_{\downarrow }+d_t^{\uparrow \uparrow}+
d_t^{\downarrow \downarrow }+d_t^0+d_A+2d_E+4t_{\uparrow }+
4t_{\downarrow}+3f\;,\\
m_{s} &=&n^0_{s}-\left[ d_t^{s s }+t_{\bar{s} }+2t_{\bar{s}}
+f +\frac{1}{2}\left( d_A+2d_E+d_t^0\right) \right]
\end{eqnarray}
as  functions of the  remaining nine independent parameters.
The expectation value of the two-band Hubbard Hamiltonian is then given by
\begin{eqnarray}\label{11.40a}
E^{2{\rm b}}_{\rm atom} &=&\sum_s(q^{s}_{s})^2\varepsilon_{s ,0} +
(U'-J)(d_t^{\uparrow \uparrow }+d_t^{\downarrow \downarrow}+d_t^0)
 \\[3pt]\nonumber
&& +2(U'+J)d_E +(U+J)d_A+(2U+4U'-2J)(t_{\uparrow}+t_{\downarrow }+f)\;,
\end{eqnarray}
where we introduced the orbital-independent
 elements 
  \begin{eqnarray}\label{11.50a}
q^{s}_{s}
&=&
\frac{1}{\sqrt{n^0_{s}(1-n^0_{s})}}
\biggl[ \left( 
\sqrt{t_{s} }+\sqrt{m_{\bar{s}}}\right) 
\frac{1}{2}\left( \sqrt{d_A}+2\sqrt{d_E}+\sqrt{d_t^0}\right)  
\\[3pt]\nonumber 
&&\hphantom{\frac{1}{n_{s}^0(1-n_{s}^0)} \Bigl[ }
+\sqrt{m_{s} }
\left( \sqrt{m_{\emptyset}}+\sqrt{d_t^{s s }}\right) 
+\sqrt{t_{\bar{s }}}
\biggl( \sqrt{d_t^{\bar{s }\bar{s } }}+\sqrt{f}\biggr) 
\biggr]
\end{eqnarray}  
of the diagonal renormalisation matrix and the bare band energies
\begin{equation}
\varepsilon_{s ,0}=\int_{-\infty }^{E_{\text{F},s}}
{\rm d}\varepsilon \,\varepsilon D_0(\varepsilon )\;.
\end{equation}

For comparison, we will also consider the energy 
\begin{eqnarray}\label{11.60a}
E^{(2 {\rm b})}_{\text{dens}} &=&\sum_{s} (\bar{q}^{s} _{s} )^2
\varepsilon_{s ,0}+(U'-J)(d_1^{\uparrow \uparrow}
+d_1^{\downarrow \downarrow })  \\[3pt] \nonumber 
&&+2U'd_0+2Ud_c+(2U+4U'-2J)(t_{\uparrow }+t_{\downarrow}+f)\;
\end{eqnarray}
of a two-band model without the terms in the 
second line of the atomic Hamiltonian~(\ref{2.160a}) since
 this is an approximation that is often made in studies on multi-band models. 
In this case, there are seven variational parameters  $d_1^{\uparrow \uparrow }$, 
$d_1^{\downarrow \downarrow }$, $d_0$, $d_c$, $t_{\uparrow }$, $t_{\downarrow}$,
and $f$, which represent the occupation of the configuration states $\ket{I}$. 
The probabilities for an empty site $m_{\emptyset}$ and a singly-occupied
site $m_{s} $ are related to the variational parameters by 
\begin{eqnarray}\label{11.80a} 
m_{s} &=&n^0_{s}-\left[ d_1^{s s }+t_{\bar{s} }+2t_{s}
+f+d_c+d_0\right] \;,  \\[3pt]
m_{\emptyset} &=&1-2n^0_{\uparrow }-2n^0_{\downarrow }+d_1^{\uparrow \uparrow}
+d_1^{\downarrow \downarrow }+2d_0+2d_c+4t_{\uparrow }+4t_{\downarrow}+3f\;.
\end{eqnarray}
The  renormalisation factors have the form
\begin{eqnarray}\label{11.70a} 
\bar{q}^s_{s} &=&
\frac{1}{\sqrt{n^0_{s}(1-n^0_{s})}}
\biggl[ 
\left( \sqrt{t_{s}}+\sqrt{m_{\bar{s} }}\right) 
\left( \sqrt{d_c}+\sqrt{d_0}\right)  
 \\[3pt]\nonumber
&&\hphantom{\frac{1}{n_{\sigma}^0(1-n_{\sigma}^0)} \Bigl[ }
+\sqrt{m_{s} }\left( \sqrt{m_{\emptyset}}+\sqrt{d_1^{s s }}\right) 
+\sqrt{t_{\bar{s} }}\biggr( \sqrt{d_1^{\bar{s}\bar{s} }}+\sqrt{f}\biggr) 
\biggr]\;.
\end{eqnarray}
\vspace{0.3cm}
{\bf C) Ground-State Properties}
\vspace{0.1cm}

\begin{figure}[tt]
\includegraphics[clip,width=7.4cm]{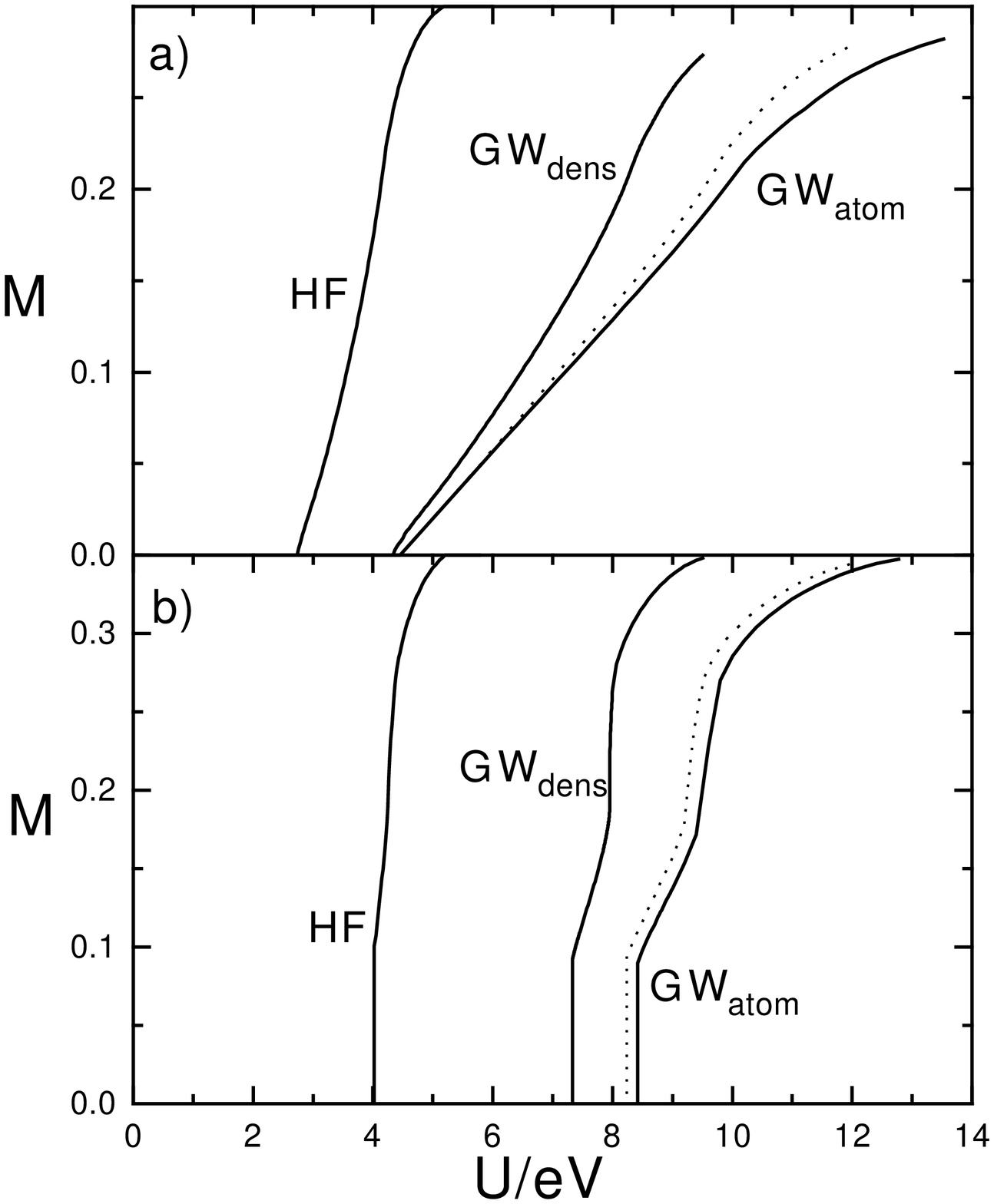}
\includegraphics[clip,width=7.4cm]{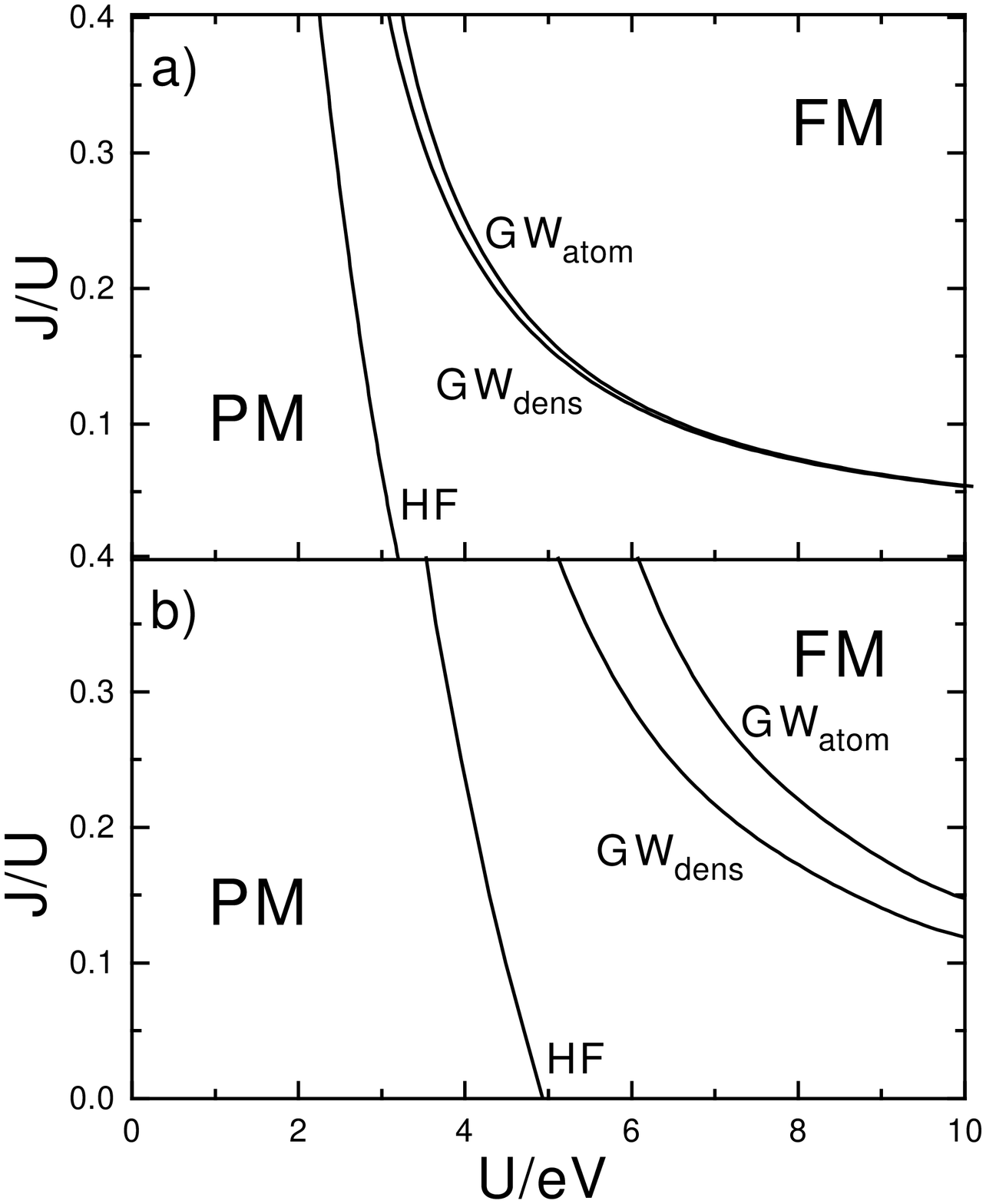}
\caption{Left:  Magnetisation density per band as a function of $U$ 
 for the Hartree--Fock solution (HF), the Gutzwiller wave function with
pure density correlations (GW$_{\text{dens}}$), and the Gutzwiller wave function
with atomic correlations (GW$_{\text{atom}}$) 
for (a) $n_s=0.29$ and (b) $n_s=0.35$.
The dotted line indicates 
the results for GW$_{\text{atom}}$ with $J_{\text{C}}=0$. The local 
 exchange interaction is $J=0.2U$ in all curves.
Right: Phase diagram as a function of $U$ and $J$ for the Hartree--Fock
solution (HF) and the two Gutzwiller wave functions
(GW$_{\text{dens}}$, GW$_{\text{atom}}$) 
for (a) $n^0=0.29$ and (b) $n^0=0.35$; PM: paramagnet, FM: ferromagnet.}\label{fig10.2ges}
\end{figure}

The energies~(\ref{11.40a}) and~(\ref{11.60a}) 
have to be minimised with respect to their respective 
 (nine or seven) independent variational parameters $m_{\Gamma}$ 
and the magnetisation
\begin{equation}\label{11.90} 
M\equiv(n^0_{\uparrow }-n^0_{\downarrow })/2\;,
\end{equation}
for example, by means of the algorithm introduced in Ref.~\cite{buenemann2012b}.
In Fig.~\ref{fig10.2ges}~(left), the magnetisation $M$ is shown 
as a function of $U$ for fixed $J/U=0.2$ ($U'/U=0.6$). 
The critical interaction for the ferromagnetic transition, 
$U_{\text{F}}^{\text{atom}}$, is
about a factor two larger than its value 
$U_{\text{F}}^{\text{HF}}$ as obtained from
the Hartree--Fock--Stoner theory. The corresponding 
values $U_{\text{F}}^{\text{dens}}$ always
lie somewhat below the values for the
Gutzwiller wave function with full atomic correlations.
In general, the relation $M_{\text{HF}}(U)> 
M_{\text{dens}}(U)> M_{\text{atom}}(U)$ holds, i.e.,
for all interaction strengths, 
the tendency to ferromagnetism is the strongest within the Hartree--Fock
theory and weakest for Gutzwiller wave functions with atomic correlations.
Furthermore, the slopes of $M(U)$ are
much steeper in the Hartree--Fock results than in the presence of
correlations. 

\begin{SCfigure}[1]\label{fig10.3}
{\centering
\includegraphics[clip,width=8cm]{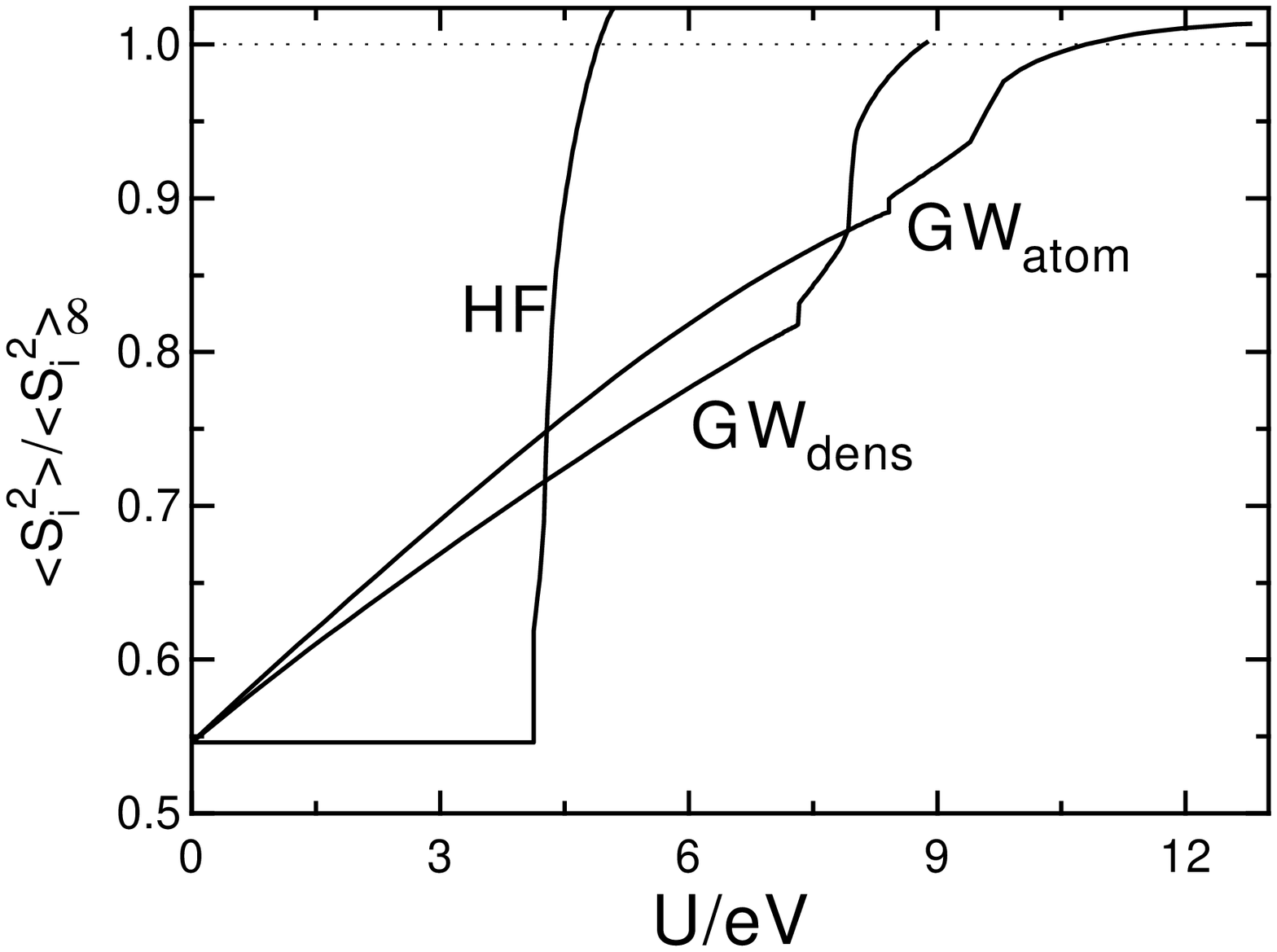}
\caption{Size of the local spin $\langle \bigl(\hat{\vec{S}}_i\bigr)^2
\rangle $ as a function of the interaction strength for $J=0.2U$ and
band filling $n^0=0.35$ for the Hartree--Fock theory (HF) and the Gutzwiller
wave functions (GW$_{\text{dens}}$, GW$_{\text{atom}}$).}}
\end{SCfigure} 

The properties of the ferromagnetic phase
strongly depend on the spectrum of the atomic two-electron
configurations. To further analyse this point, we have included
the case of $J_{\text{C}}=0$, which changes only 
the excited two-electron states. A shift of the curve $M(U)$ results
towards smaller interaction strengths; for a given
magnetisation density, a smaller interaction strength is required as
compared to the correct symmetry case $J=J_{\text{C}}$, 
see Fig.~\ref{fig10.2ges}~(left).
The effect is more pronounced when we go to the Gutzwiller wave function
with pure density correlations. 
These results indicate that itinerant ferromagnetism
is strongly influenced by the atomic multiplet spectra.

In Fig.~\ref{fig10.2ges}~(left/a),
we chose the particle density per band to be
$n^0=(n^0_{\uparrow}+n^0_{\downarrow})/2=0.29$, right at
the maximum of the density of state curve, compare Fig.~\ref{fig10.1}.
In this case, there are finite slopes of the $M(U)$ curves at $U_{\text{F}}$, 
and a `Stoner criterion' for the onset of ferromagnetism 
applies. In Fig.~\ref{fig10.2ges}~(left/b), we chose
the particle density per band as $n^0=0.35$. 
In this case, the density of states at the Fermi energy 
$D_0(E_{\text{F},\uparrow})+D_0(E_{\text{F},\downarrow })$ 
first {\em increases\/} as a function of the magnetisation density.
 Therefore, a discontinuous transition thus occurs from the paramagnet to
the ferromagnet.

In the case of pure density correlations, a second jump in the 
$M(U)$ curve is observed
 that is absent in the other two curves. 
As discussed in Ref.~\cite{buenemann1997b}, 
this jump is related to another feature of the density of states. In the
Hartree--Fock theory, this feature is too weak to be of any significance 
in comparison to the interaction energy. When the full 
atomic correlations are taken into account, this 
first-order jump at a finite magnetisation density
disappears due to the enhanced flexibility of the 
variational wave function.

Another remarkable difference between the Hartree--Fock and the Gutzwiller
method lies in the approach to ferromagnetic
saturation.
In the Hartree--Fock theory, the magnetisation saturates
at $U$~values about 20\% to 40\% above
the onset of ferromagnetism at $U_{\text{F}}^{\text{HF}}$. 
In contrast, in the variational approach saturation is reached
at about twice the onset value, $U_{\text{sat}}\lesssim 2U_{\text{F}}$. 
However, even when the minority spin occupancies are zero
and $\langle \hat{S}_z^{\text{at}} \rangle$ is constant, the majority spin
occupancies $s_{\uparrow }$ and $d_t^{\uparrow \uparrow }$ vary 
with $U$ since the limit of zero empty sites is reached
only for $U\to\infty$.

In Fig.~\ref{fig10.2ges}~(right), we display the $J$-$U$ phase diagram for both
fillings. It shows that the Hartree--Fock theory always predicts a ferromagnetic
instability. In contrast, the correlated electron approach strongly supports
the idea that a substantial on-site Hund's rule exchange is 
required for the occurrence
of ferromagnetism at realistic interaction strengths. For the case
$n^0=0.29$, the differences between the phase diagrams for the two
correlated electron wave functions are minor. Due to the large density
of states at the Fermi energy, the critical interaction strengths 
for the ferromagnetic transition are comparably small, 
and the densities for the double occupancies in both
correlated wave functions do not differ much. 
For the larger band filling $n^0=0.35$, i.e., away from 
the peak in the density of state, 
the values for $U_{\text{F}}$ are considerably larger and, in the
atomic correlation case, the Gutzwiller wave functions can generate
local spin triplets  more easily while keeping the global paramagnetic phase.

The magnitude of the local spin as a function of $U$ is shown
in Fig.~\ref{fig10.3}.
For $U\to\infty$, each site is either singly 
occupied with probability $2-4n^0$ or doubly occupied (spin $S=1$) 
with probability $4n^0-1$. Hence, 
\begin{equation}\label{11.100} 
\langle (\vec{S}_i)^2\rangle_{\infty} =(3/4)(2-4n^0)+2(4n^0-1)=
5n^0-1/2\;.
\end{equation}
For the correlated wave functions, 
this limit is reached from {\em above\/} since, for $U<\infty$, charge
fluctuations first increase the number of spin-one sites at the expense of
spin-$1/2$ sites, which turn into empty sites. A further decrease of $U$ will
also activate the singlet double occupancies and higher multiple
occupancies. Thus, the local spin eventually reduces
below $\langle (\vec{S}_i)^2\rangle_{\infty}$. 
On the contrary, the Hartree--Fock theory
does not give the proper large-$U$ limit for the local spin.
Instead, the Hartree--Fock limit is
given by 
\begin{equation}\label{11.110} 
\langle (\vec{S}_i)^2\rangle_{\infty}^{\text{HF}}
=n^0(3+2n^0)\;.
\end{equation}
The change of $\langle (\vec{S}_i)^2\rangle $ 
at $U_{\text{F}}$ is only a minor effect within the 
correlated electron approach. In particular, this holds true
for the case of atomic correlations, where about 90\% of the local spin 
saturation value is already reached in the paramagnetic state.
Again, the Hartree--Fock results are completely different.
There, the local spin sharply increases as a function of
the interaction strength since the absence of correlations fixes
\begin{equation}\label{11.120} 
\langle (\vec{S}_i)^2\rangle^{\text{HF}}(U< U_{\text{F}}^{\text{HF}})
= \langle (\vec{S}_i)^2\rangle(U=0).
\end{equation}

\begin{SCfigure}\label{fig10.5}
{\centering
\includegraphics[clip,width=8cm]{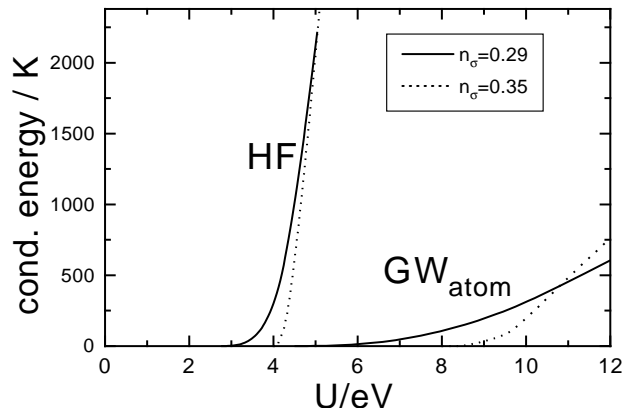}
\caption{Condensation energy as a function of $U$ for $J=0.2U$ for the
Hartree--Fock theory (HF) and the Gutzwiller wave function 
(GW$_{\text{atom}}$) for $n=0.29$ (full lines) 
and $n=0.35$ (dashed lines).}}
\end{SCfigure} 

Finally, in Fig.~\ref{fig10.5}, we display the energy differences
between the paramagnetic and ferromagnetic ground states as a function of
the interaction strength for $J=0.1U$. For the correlated electron case,
this quantity is of the order of the Curie temperature,  which is in the
range of $100\,\text{K}-1000\,\text{K}$ in real materials. 
On the other hand, the
Hartree--Fock theory yields small condensation energies only in the range 
of $U\approx 4\,$eV; for larger $U$, the condensation energy 
is of order $U$.
Including the correlation effects within the Gutzwiller theory, we have
relatively small condensation energies even for interaction values
as large as twice the bandwidth ($U\approx10\,$eV).

\subsection{Antiferromagnetic order in iron-pnictide models}\label{chap10.1b} 
Since their recent discovery, the iron-based high-$T_{\rm c}$ 
superconductors, e.g., ${\rm La O Fe As}$, have attracted tremendous 
attention both in
theory and experiment. From a theoretical point of view,
these systems are of particular interest because
their conduction electrons are less correlated 
than those of other high-$T_{\rm c}$ superconductors.
In contrast to the cuprates, the pnictides' undoped parent compounds
are antiferromagnetic metals at low temperatures,
not insulators. However, the electronic mass is enhanced by
a factor of two which indicates that electronic correlations
are quite substantial in the pnictides, too.

The theoretical description of the pnictides' normal phases already 
turned out to
be a difficult problem. Standard density-functional theory (DFT)
grossly overestimates the size of their magnetic moment
in the antiferromagnetic ground state.
For example, in ${\rm La O Fe As}$ experiment finds 
a staggered moment of 
$m= (0.4\ldots 0.8)\mu_{\rm B}$~\cite{cruz2008,qureshi2010,li2010} 
whereas DFT calculations predict moments of 
$m\approx 1.8\mu_{\rm B}$, or larger~\cite{mazin2008,zhang2010}. 
For other pnictide compounds, the comparison is equally unfavourable.

\begin{figure}[ttt]
\includegraphics[clip,width=5.1cm]{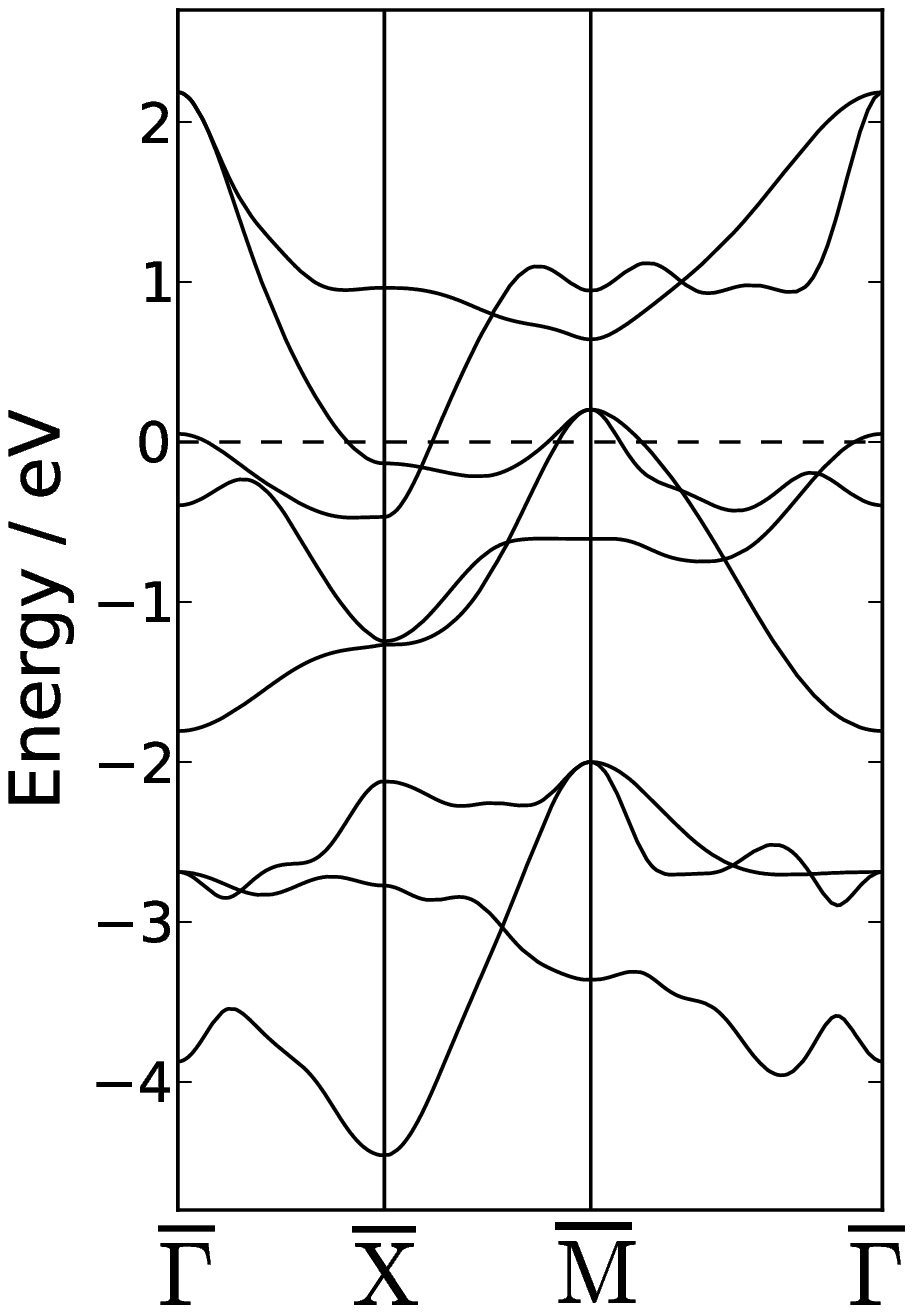}
\includegraphics[clip,width=5.1cm]{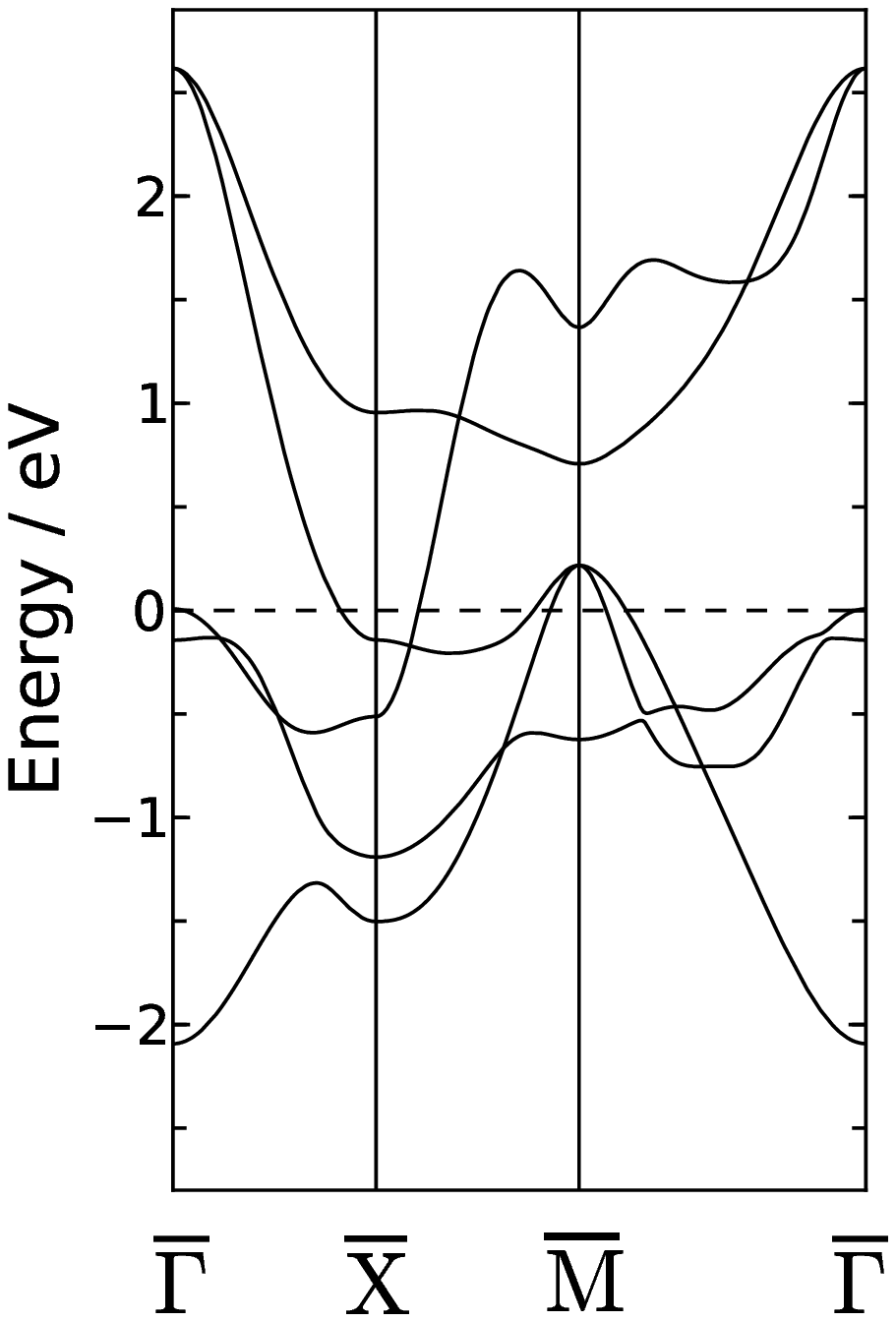}
\includegraphics[clip,width=5.1cm]{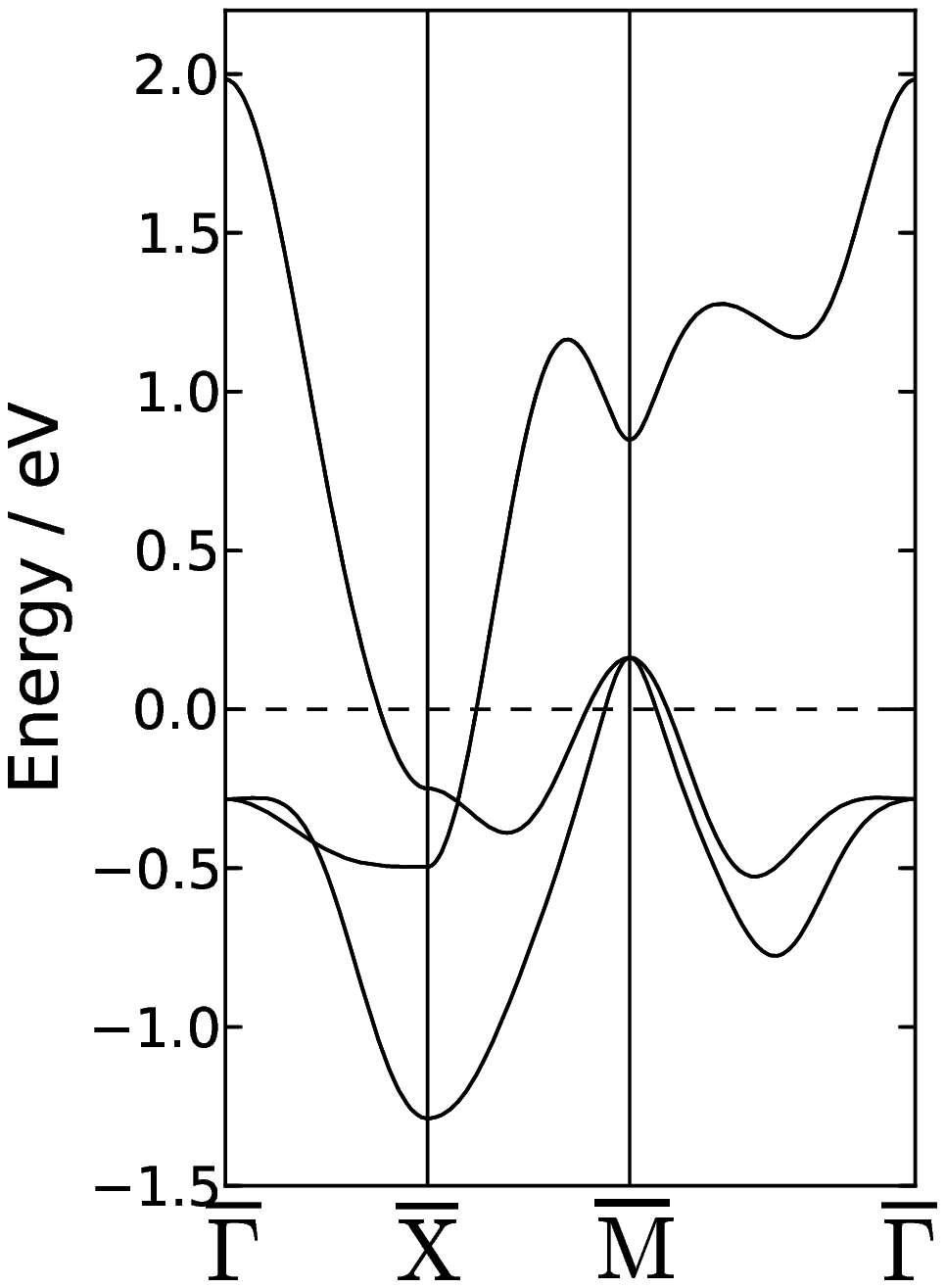}
\caption{Model band structures for  ${\rm La O Fe As}$ with eight 
bands~\cite{andersen2010}~(left), 
 five bands~\cite{graser2009}~(middle) and three bands~\cite{zhou2010} (right).
  }\label{fig789}
\end{figure}

The electronic structure of ${\rm La O Fe As}$  near the Fermi energy is 
fairly two-dimensional and the bands are dominantly of iron $d$ and 
 (partially) of arsenic $p$ character. A complete tight-binding model 
for ${\rm La O Fe As}$ should therefore consist of eight bands
(i.e., five iron $d$ and three arsenic $p$ bands)~\cite{andersen2010}, 
see Fig.~\ref{fig789}~(left). 
For many-particle approaches, however, the study of such 
 an eight-band model model is obviously quite challenging. 
 Therefore, in many theoretical works 
 on iron pnictides various simpler models have been 
 proposed to study particular aspects of these materials. The fact that
 the bands near the Fermi energy are dominantly of iron $d$   character 
  suggest the study of an effective  five-band model 
of pure iron $d$-bands. Such a model has been proposed, e.g., in 
 Ref.~\cite{graser2009}, see Fig.~\ref{fig789}~(middle). 
Even simpler models may be derived if one only aims
 to reproduce  the Fermi surfaces of 
${\rm La O Fe As}$. This is achieved, e.g., by the three-band model in 
 Fig.~\ref{fig789}~(right),  
 which was investigated in Ref.~\cite{zhou2010}.

In cases where a simplified model reproduces certain properties of a material 
 correctly, there will often remain doubts whether this agreement is merely
 coincidental or an indication that a model indeed captures the relevant 
physics of a system. A big advantage of the Gutzwiller 
theory is its numerical simplicity that allows one to study even complicated 
 multi-band models with modest numerical efforts. 
In this way, it is possible to test the quality
 of simplified models by comparing their properties with those of more realistic
 Hamiltonians.
 In this section, we will compare the magnetic properties 
of all three models, displayed in Figs.~\ref{fig789}. This 
 comparison provides an interesting example of the dangers 
 that lie in the study of `oversimplified' model systems. 
 Based on a Gutzwiller theory calculation it has been argued in 
 Ref.~\cite{zhou2010} that
 the three-band model has a relatively small magnetic moment, 
 in agreement with experiment. However, as we have shown in 
Ref.~\cite{buenemann2011}
  the magnetic properties of the five-band model are 
 very different from experiment and from those of the 
 three-band model. One must therefore conclude that  
 both models are insufficient in describing the magnetic 
 properties of ${\rm La O Fe As}$. In fact, it turns out that 
an inclusion of the arsenic $p$ orbitals is essential, see below. 
 
 \begin{SCfigure}[1]\label{figg1}
 \centering
 \includegraphics[width=0.5\textwidth]{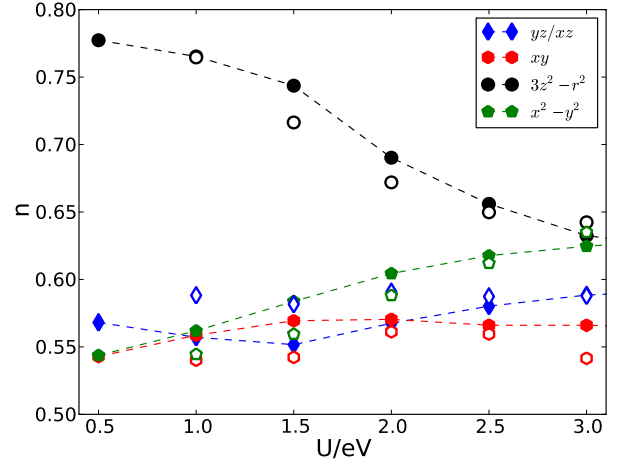}
 \caption{
Orbital densities in Gutzwiller theory (full symbols)
and in DMFT (open symbols)
as a function of $U$ (with $U/J=4$) for the simplified local Hamiltonian
$\hat{H}_{\rm c}=\hat{H}_{\rm c}^{(1)}$, see Eq.~(\ref{app3.5}).
}
\end{SCfigure} 

In many theoretical studies,
 the following Hamiltonian for the Hubbard interaction 
$\hat{H}_{i;{\rm c}}$ in~(\ref{er})  is used,
\begin{eqnarray}\label{app3.5}
\hat{H}^{(1)}_{\rm c}&=&\hat{H}^{\rm dens}_{\rm c}+\hat{H}^{\rm sf}_{\rm c} \;,\\
\hat{H}^{\rm dens}_{\rm c}&=&
\sum_{b,s}U(b,b)
\hat{n}_{b,s}\hat{n}_{b',\bar{s}}
+\sum_{b(\neq)b'}\sum_{s,s'}
\widetilde{U}_{s,s'}(b,b')
\hat{n}_{b,s}\hat{n}_{b',s'} \, ,\nonumber\\
\hat{H}^{\rm sf}_{\rm c} &=&
\sum_{b(\neq)b'}J(b,b')\left(\hcd_{b,\uparrow}\hcd_{b,\downarrow}
\hc_{b',\downarrow}\hc_{b',\uparrow}+ {\rm h.c.}\right) 
+\sum_{b(\neq)b';s}J(b,b')\hcd_{b,s}\hcd_{b',\bar{s}}
\hc_{b,\bar{s}}\hc_{b',s}\;.
\nonumber
\end{eqnarray}
Here, we introduced
 $\widetilde{U}_{s,s'}(b,b')= U(b,b')-\delta_{s,s'}J(b,b')$,
where $U(b,b')$ and  $J(b,b')$ are the local Coulomb and exchange 
interactions. 
For a system of five correlated $d$ orbitals in cubic environment, however, the 
Hamiltonian~(\ref{app3.5})
is incomplete~\cite{sugano1970}. The full Hamiltonian reads 
$\hat{H}_{\rm c}=\hat{H}^{(1)}_{\rm c}+\hat{H}^{(2)}_{\rm c}$ where
\begin{eqnarray}\nonumber
\hat{H}^{(2)}_{\rm c}&=&\bigg[\sum_{t; s,s'}(
T(t)-\delta_{s,s'}A(t))
\hat{n}_{t,s}\hcd_{u,s'}\hc_{v,s'}+\sum_{t,s}A(t)
\left(
\hcd_{t,s}\hcd_{t,\bar{s}}
\hc_{u,\bar{s}}\hc_{v,s}+
\hcd_{t,s}\hcd_{u,\bar{s}}
\hc_{t,\bar{s}}\hc_{v,s}
\right)\\
\label{h255}
&&+\sum_{ t(\neq)t'(\neq)t^{\prime \prime}}
\sum_{e,s,s'}
S(t,t';t^{\prime \prime},e)
\hcd_{t,s}\hcd_{t',s'}
\hc_{t^{\prime \prime},s'}\hc_{e,s}\bigg ]+{\rm h.c.}\,.
\end{eqnarray}
Here, $t$ and $e$ are indices for the  
three $t_{2g}$ orbitals with symmetries $xy$, $xz$, and $yz$,
and the two $e_g$ orbitals with symmetries
$u=3z^2-r^2$ and $v=x^2-y^2$. 
The parameters in~(\ref{h255}) are of the same order of magnitude 
as the exchange interactions $J(b,b')$ and, hence, 
there is no a-priori reason to neglect them. 
Of all the parameters $U(b,b')$, $J(b,b')$,
$A(t)$, $T(t)$, $S(t,t';t^{\prime \prime},e)$ 
only ten are independent in cubic symmetry. 
In a `spherical approximation', i.e., assuming that all orbitals have 
 the same radial wave-function, all parameters are determined by, e.g., the 
three Racah parameters $A,B,C$.  We prefer to work with the orbital averages 
$J\propto\sum_{b\neq b'}J(b,b')$, 
and $U'\propto\sum_{b\neq b'}U(b,b')$ 
of the exchange and the inter-orbital Coulomb interaction. 
They are related to the intra-orbital interaction 
$U=U(b,b)$ via $U'=U-2J$ . 
Due to this symmetry relation, the three values of $U,U',$ and $J$ do not 
determine the Racah parameters $A,B,C$ uniquely. Therefore, we make use
 of the atomic relation $C/B=4$ which is approximately satisfied in 
solids, too. In this way, the three Racah parameters and, 
consequently, all parameters 
in $\hat{H}_{\rm c}$ are functions of $U$ and~$J$.
This permits a meaningful comparison of our results for all three 
 model Hamiltonians.
 
In order to test the reliability of our approach  
we first compare our results for the partial densities of the five-band model 
with those from paramagnetic DMFT calculations. In Fig.~\ref{figg1} 
we show the density of electrons in each orbital 
as a function of~$U$ for fixed ratio  $U/J=4$.
The full symbols give the Gutzwiller result 
for the simplified local Hamiltonian~(\ref{app3.5}),
$\hat{H}=\hat{H}_0+\hat{H}_{\rm c}^{(1)}$;
open symbols give the DMFT results ~\cite{ishida2010}.
Obviously, the agreement between the Gutzwiller theory and DMFT is very good.
 This comes not as a surprise because both methods are 
 derived in the limit of infinite spatial dimensions.

Figure~\ref{figg1} shows a common feature of multi-band model systems.
The local Coulomb interaction induces a substantial charge flow between
the bands because, for the local Coulomb interaction, it is energetically
more favourable to distribute electrons equally among the bands. However,
the bands described by $\hat{H}_0$ are extracted from a DFT calculation
whose predictions for the Fermi surface reproduce experimental data
reasonably well. Therefore, the artificial charge flow 
as seen in Fig.~\ref{figg1} is clearly a consequence
of the double counting of Coulomb interactions.
Since the (paramagnetic) Fermi surface found in DFT
reproduces its experimentally determined shape, 
we assume that the same holds for the paramagnetic 
orbital densities. For each value of the interaction parameters we therefore 
choose orbital on-site energies $\epsilon_{i}^{\sigma,\sigma}$ 
which lead to a paramagnetic ground state with the same orbital densities 
as in DFT. Note that a more sophisticate calculation of orbital densities 
 requires the self-consistent Gutzwiller DFT scheme which we shall 
 introduce in the following section. 

\begin{figure}[ttt]
\includegraphics[clip,width=5.1cm]{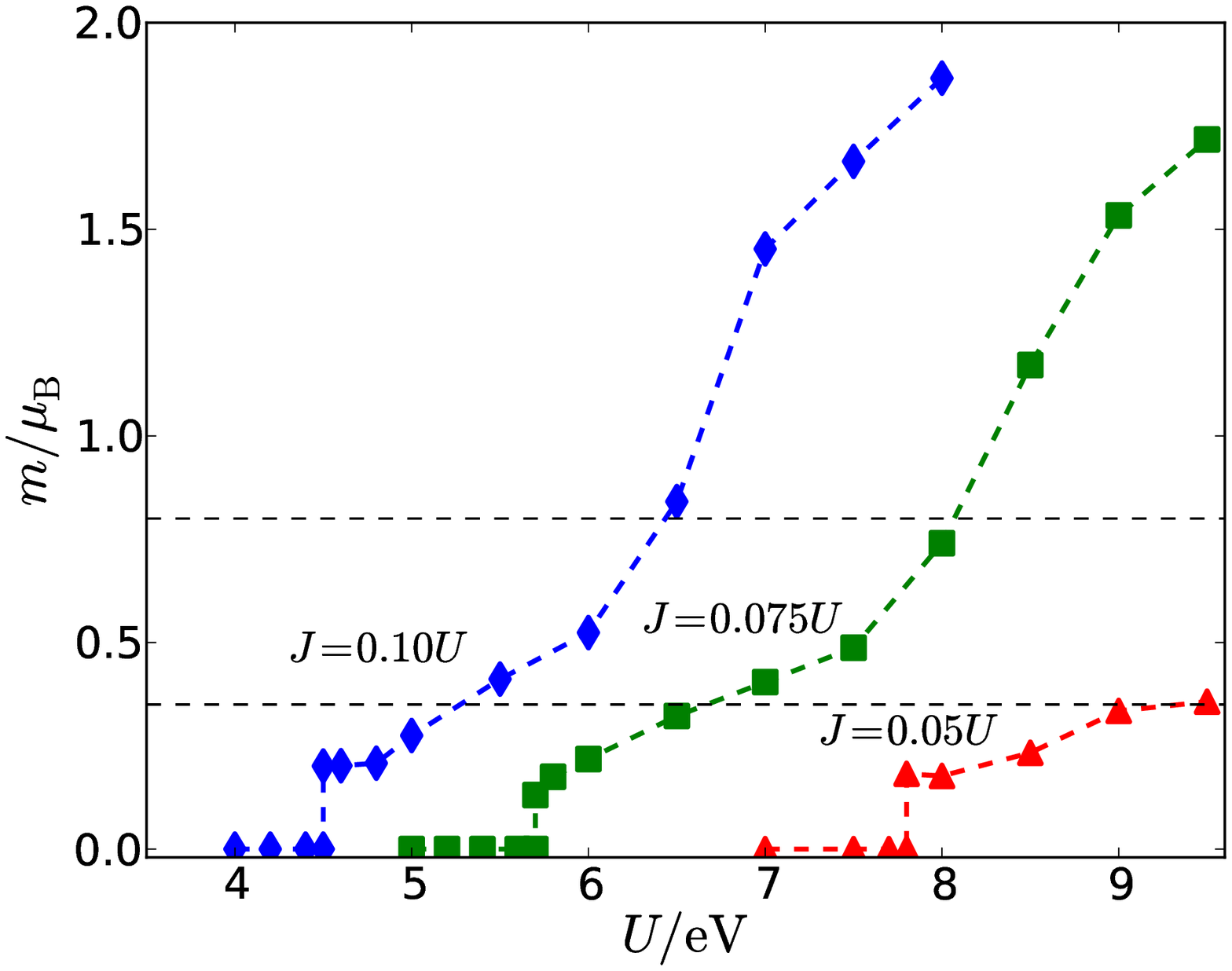}
\includegraphics[clip,width=5.1cm]{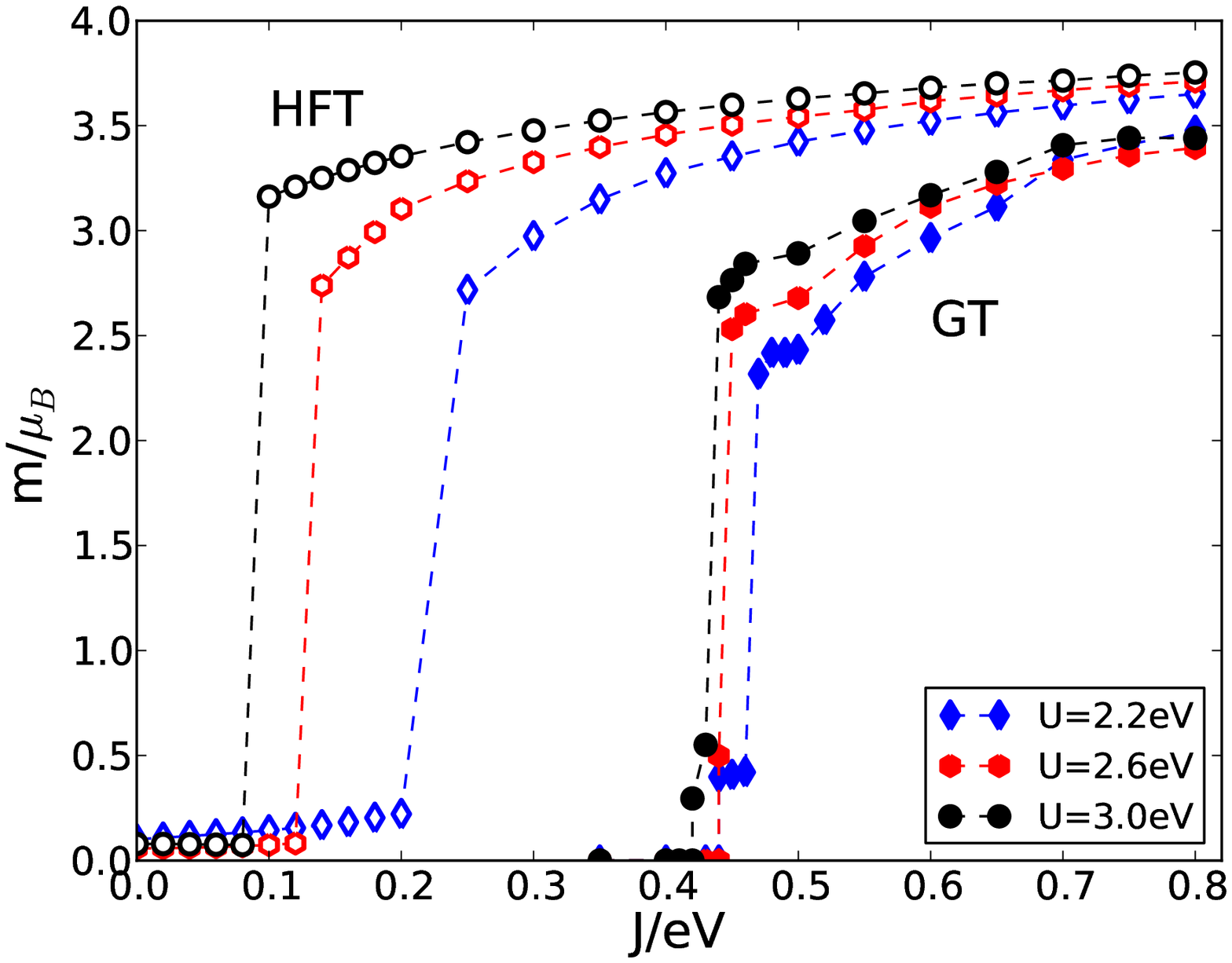}
\includegraphics[clip,width=5.1cm]{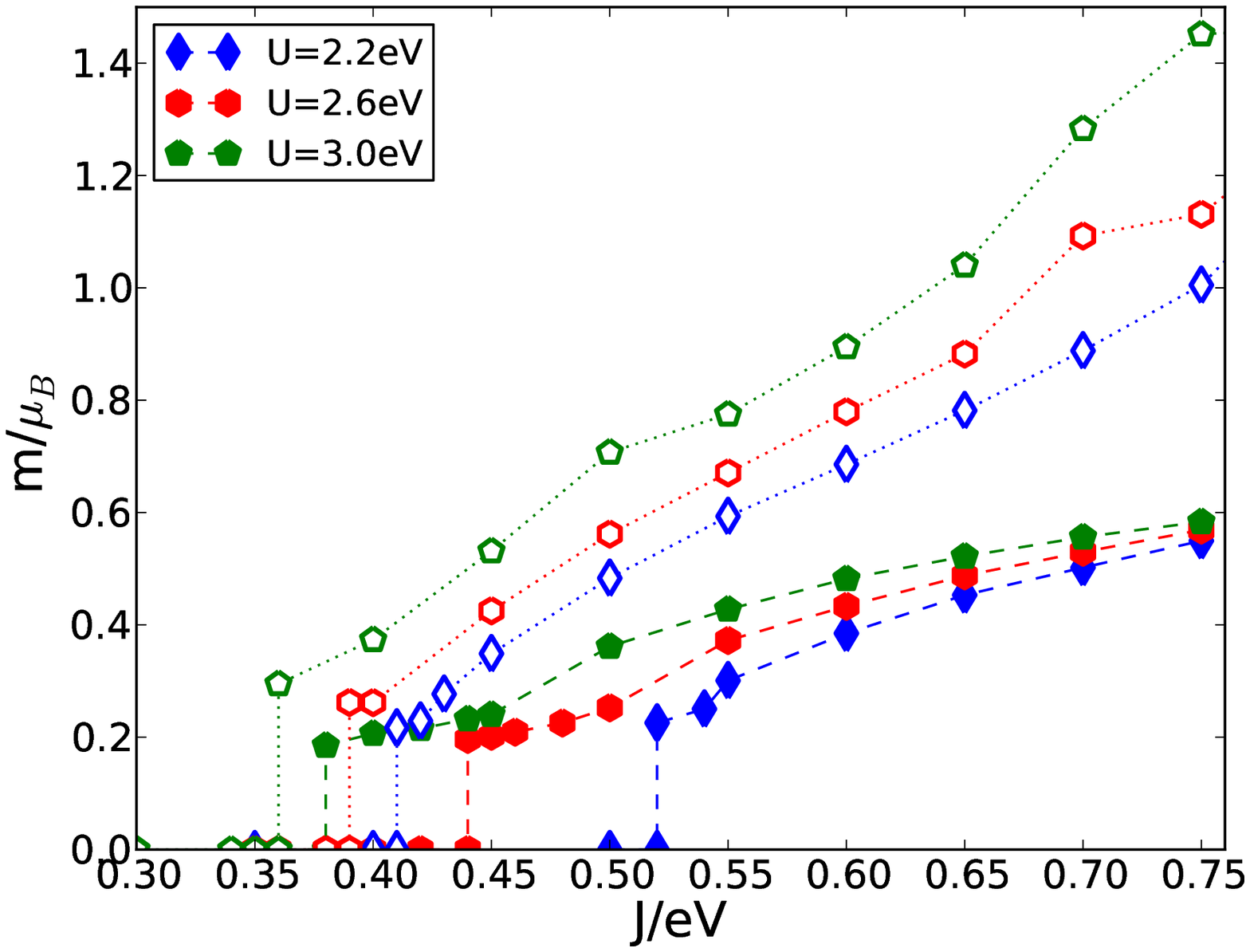}
\caption{Magnetic moment as a function of $U$ for: left: model with eight 
bands~\cite{andersen2010}; middle: 
 model with five bands~\cite{graser2009} (Hartree--Fock and Gutzwiller theory); 
 right: model with three bands~\cite{zhou2010} and local Hamiltonians 
$\hat{H}^{\rm dens}_{\rm c}$ (dotted) and $\hat{H}^{\rm sf}_{\rm c}$ (dashed) .
  }\label{fig789b}
\end{figure} 

In Figs.~\ref{fig789b} we display the magnetic moment as a function of $U$ for all  
 three model systems. As mentioned before, the  three-band model shows relatively small magnetic moments over a large range of Coulomb- and 
 exchange interaction parameters, see Fig.~\ref{fig789b}~(right).
 This is in stark contrast to the results for the
  five-band model in Fig.~\ref{fig789b}~(middle). Here we find a
 transition to an antiferromagnet 
with a large moment,
$m\gtrsim 2\mu_{\rm B}$ which is as abrupt as in a corresponding
  Hartree--Fock calculation.   
However, if one takes into account
 the arsenic $p$ bands, the magnetic moment is significantly smaller
 and in the range of experimental without the need of `fine-tuning' the 
 Coulomb- or exchange interaction parameters, see Fig.~\ref{fig789b}~(left). 
In summary, we can conclude
 that a proper understanding of the magnetic order in  ${\rm La O Fe As}$
 requires the study of an eight-band model of iron $d$ and arsenic $p$
 bands. It it possible, of course, that other properties of this compound, 
 e.g., the superconducting order, may be correctly described by simpler 
 model Hamiltonians. 
\section{The Gutzwiller density-functional theory}\label{gdft} 
\subsection{The Gutzwiller DFT equations}
 The model-based Gutzwiller method, which we have used in the previous
 section, 
ignores the fact that the hopping parameters~(\ref{tzr})  
are actually functions of the density $n(\ve{r})$. Taking this functional 
dependence into account defines the `Gutzwiller density functional theory' (GDFT). 
The explicit inclusion of the local Coulomb interaction within the 
Gutzwiller theory leads to changes of the 
particle density $n(\ve{r})$ for three reasons:
\begin{itemize}
\item[i)]
The particle density $n(\ve{r})$ in the Gutzwiller-correlated ground state
\begin{equation}\label{aw}
n(\ve{r})=\sum_{i\ne j} \sum_{\sigma,\sigma',\gamma,\gamma'}
\phi^*_{i,\gamma}(\ve{r})
\phi_{j,\gamma'}(\ve{r})q^{\sigma}_{\gamma} 
\left( q^{\sigma'}_{\gamma'}\right)^{*} 
\rho_{(j\sigma'),(i\sigma)}+
\sum_{i} \sum_{\sigma}|\phi_{i,\sigma}(\ve{r})|^2\rho_{(i\sigma),(i\sigma)}
\end{equation}
differs from the corresponding DFT expression~(\ref{pden}).
\item[ii)]
Due to the renormalisation factors in~(\ref{1.4c})
there will be an energy gain from changes of the hopping parameters 
$t^{\sigma,\sigma'}_{i,j}=t^{\sigma,\sigma'}_{i,j}\!\!\left\{n(\ve{r})\right\}$, which  
requires the re-adjustment of $n(\ve{r})$.
\item[iii)]
The Coulomb interaction can lead to drastic changes of the 
occupation numbers $n^0_{i\sigma}=\rho_{(i\sigma),(i\sigma)}$ in the 
localised orbitals, e.g., when the ground state is magnetically ordered. 
This also changes the non-local elements of 
the single-particle density matrix
$\tilde{\rho}$ and the particle density~(\ref{aw}).
\end{itemize}

These correlation-induced changes of the particle density 
are taken into account in the self-consistent 
GDFT by including the dependence of the hopping parameters on $n(\ve{r})$.
Equation~(\ref{aw}) shows that $n(\ve{r})$ and, consequently,  
$t^{\sigma,\sigma'}_{i,j}$ are unique functions of $\tilde{\rho}$ and 
 $\lambda_{\Gamma}$. 
Therefore, the GDFT energy functional has the form
\begin{equation}  
\label{123}
E^{\rm GDFT}\left(\tilde{\rho},\lambda_{\Gamma}\right) =\!\!
\sum_{\sigma,\sigma',\gamma,\gamma'}\!\!
q^{\sigma}_{\gamma} \left( q^{\sigma'}_{\gamma'}\right)^{*} \!
\sum_{i \ne j } t^{\gamma,\gamma'}_{i,j}(\tilde{\rho},\lambda_{\Gamma}) 
\rho_{(j\sigma'),(i\sigma)}
+\sum_{i , \sigma \in {\rm d}} \epsilon_i^{\sigma,\sigma}\rho_{(i\sigma),(i\sigma)}
+ L\sum_{\Gamma}E_{\Gamma}\lambda^{2}_{\Gamma}m^0_{\Gamma}  \; .
\end{equation} 
We assume again that the constraints~~(\ref{5.5}),~(\ref{5.5b}) 
are solved by expressing some of the parameters $\lambda_\Gamma$ by 
 the remaining `independent' parameters $\lambda^{\rm i}_\Gamma$.
The resulting energy functional 
$\bar{E}^{\rm GDFT}\left(\tilde{\rho},\lambda^{\rm i}_{\Gamma}\right)$ has to be minimised with respect to 
the density matrix $\tilde{\rho}$ 
and the independent variational parameters $\lambda^{\rm i}_{\Gamma}$,
\begin{equation} 
\label{12}
\frac{\partial}{\partial \rho_{(i\sigma),(j\sigma')}}
\bar{E}^{\rm GDFT}\left(\tilde{\rho},\lambda^{\rm i}_{\Gamma}\right)=0\quad,\quad
\frac{\partial}{\partial \lambda^{\rm i}_{\Gamma}}
\bar{E}^{\rm GDFT}\left(\tilde{\rho},\lambda^{\rm i}_{\Gamma}\right)=0\;,
\end{equation} 
with the usual constraint~(\ref{16}) for the non-interacting 
density matrix $\tilde{\rho}$.  The minimisation 
with respect to $\tilde{\rho}$ leads to 
`renormalised' Kohn--Sham equations of the form~(\ref{er1}), 
(\ref{tzr}), (\ref{er4}), and (\ref{aw}) with Eq.~(\ref{er1}) replaced by 
\begin{equation}
\label{er1zz}
\hat{H}_0=\sum_{i\ne j} \sum_{\sigma,\sigma',\gamma,\gamma'}
q^{\sigma}_{\gamma} \left( q^{\sigma'}_{\gamma'}\right)^{*} 
t^{\gamma,\gamma'}_{i,j}\!\!(\tilde{\rho},\lambda_{\Gamma}) \hcd_{i,\sigma}\hc_{j,\sigma'} 
+\sum_{i,\sigma\in \ell}\eta_{\sigma}\hcd_{i,\sigma}\hc_{i,\sigma} \;
\end{equation}
and by \cite{note2}
\begin{equation}
\label{er1zz2}
\eta_{\tilde{\sigma}}\equiv
\frac{1}{L}\sum_{\sigma,\sigma',\gamma,\gamma'} 
\Big[\frac{\partial}{\partial n_{\tilde{\sigma}}}
q^{\sigma}_{\gamma} \left( q^{\sigma'}_{\gamma'}\right)^{*}\Big]
\sum_{i \ne j}t^{\gamma,\gamma'}_{i,j}\!\!(\tilde{\rho},\lambda_{\Gamma})
\rho_{(j\sigma'),(i\sigma)}
+\frac{\partial}{\partial n_{\tilde{\sigma}}}\sum_{\Gamma}              
E_{\Gamma}\lambda^{2}_{\Gamma}m^0_{\Gamma} \;,
\end{equation}
respectively. The set of Eqs.~(\ref{tzr}),  
(\ref{aw}), and (\ref{123})-(\ref{er1zz2}), which have to be solved
 self-consistently, constitute the GDFT.
This approach was first proposed in Refs.~\cite{ho2008,deng2008,deng2009} 
and has been applied to various systems in 
Refs.~\cite{deng2008,deng2009,wang2008,wang2010,weng2011}.
In all these works, the authors report a remarkably better
agreement with experiment than it could be obtained by a model-based
Gutzwiller calculation for the same materials. 
 
One of the main advantages of the DFT is its `ab-initio' character, 
i.e., the absence of any adjustable parameters. Unfortunately, this
benefit of the DFT cannot be fully maintained in the
GDFT because that would require the calculation of the 
two-particle interaction parameters  $U_{i}^{\sigma_1,\sigma_2,\sigma_3,\sigma_4}$
 in the localised orbitals from first principles.
The straightforward {\sl ab-initio\/} solution of this problem, 
namely to calculate these parameters  from 
the Wannier orbitals  $\phi_{i,\sigma}(\ve{r})$, 
is known to yield values which are much too large. 
Apparently, screening effects decrease the Coulomb-interaction
parameters significantly. These effects, however, are not well understood 
and a quantitatively reliable technique for the calculation of screened 
Coulomb parameters does not exist. 
For this reason one usually applies the same strategy as in 
model-based calculations where the matrix elements 
$U_{i}^{\sigma_1,\sigma_2,\sigma_3,\sigma_4}$ are 
somehow parameterised, e.g.,  in spherical approximation 
by means of a few Racah parameters. These are chosen to obtain the
best agreement with experiment. In this context, it is a big advantage that 
the GDFT provides one with more data, e.g., 
with structural properties, that can be compared to experiment, 
see Sec.~\ref{4rt}.  
  
As mentioned before, the local Coulomb interaction appears 
twice in the Hamiltonian~(\ref{er}) because
it also affects the on-site energies $\epsilon_i^{\sigma,\sigma'}$. 
There are several ways to overcome this `double-counting problem' 
which have been proposed in the literature, 
see, e.g., Ref.~\cite{karolak2010}. According to 
Refs.~\cite{deng2008,deng2009,wang2008,wang2010,weng2011}, 
the subtraction of the mean-field operator
\begin{equation}
\hat{H}_{\rm dc}=2\sum_{\sigma,\sigma', \gamma \in \ell}
(U_{i}^{\sigma,\gamma,\gamma,\sigma'}-U_{i}^{\gamma,\sigma,\gamma,\sigma'})
n^0_{\gamma}\hcd_{i,\sigma} \hc_{i,\sigma'}
\end{equation}
from $\hat{H}_{i;{\rm c}}$ leads to good results within the GDFT. 

\subsection{Application}\label{4rt}
\begin{figure}[tt]
 \centering
 \includegraphics[width=0.95\textwidth]{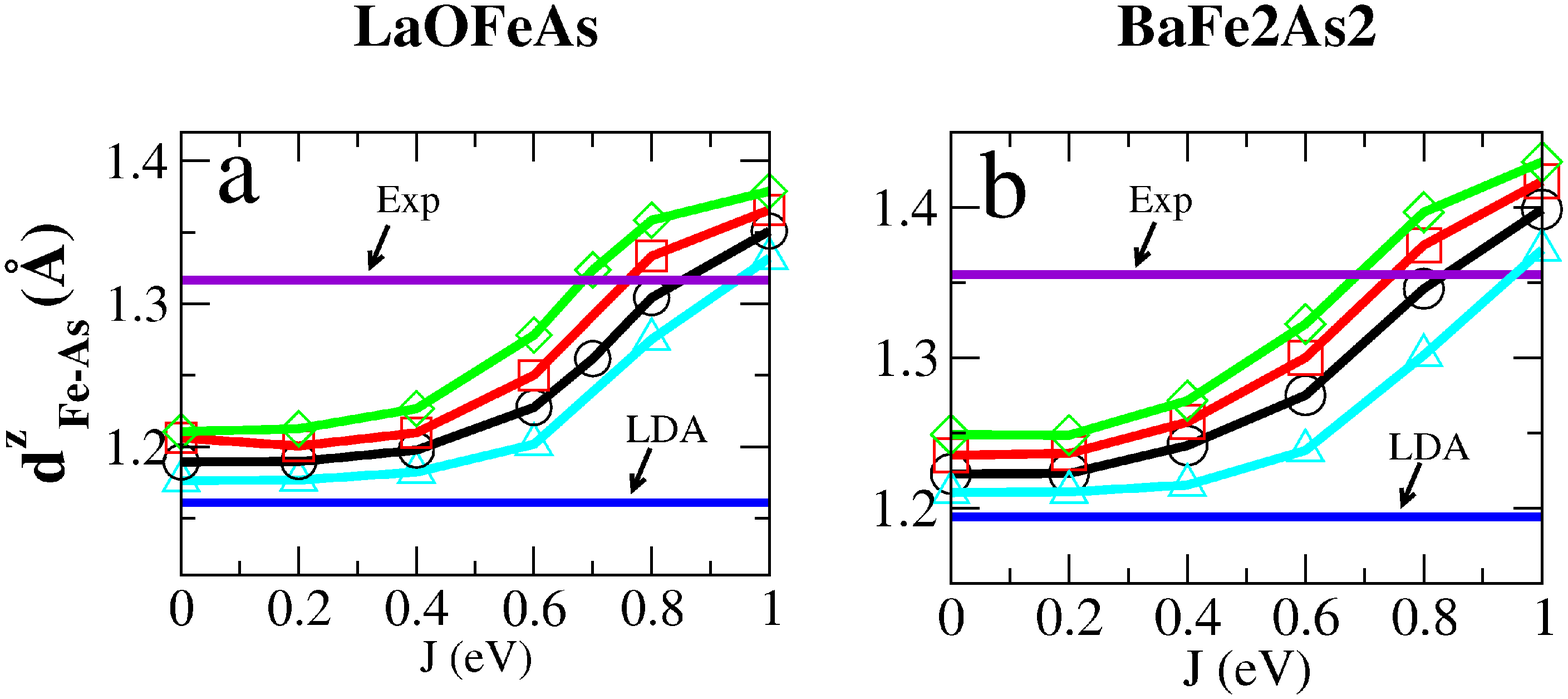}
 \caption{Interlayer distance $d^z_{\rm FeAs}$ in ${\rm La O Fe As}$ and 
  ${\rm Ba Fe_2 As_2}$ as a function of $J$ for several 
 values of $U$.}\label{fig10b}
\end{figure}

As an example for the relevance of the GDFT 
  we show results on the iron-pnictides ${\rm La O Fe As}$ and 
  ${\rm Ba Fe_2 As_2}$ which have been presented in Ref.~\cite{wang2010}. 
 The failure to describe the magnetic order  is  
 not the only problem the DFT faces in its calculations on iron-pnictides. 
 There are also substantial deviations
 between the experimental results and DFT predictions on lattice 
 parameters, in particular for the distance between ${\rm Fe}$ and ${\rm As}$. 
 Taking correlations into account more properly, as it is done within the GDFT, 
 changes these lattice parameters significantly. Figure~\ref{fig10b} shows the 
interlayer distance $d^z_{\rm FeAs}$ as a function of $J$ for several 
 values of $U$. In both systems, the exchange interaction clearly plays an 
 important role and needs to be included in order to reproduce the 
 experimental value for $d^z_{\rm FeAs}$. As a consequence, other 
 properties are also changed significantly as a function of $J$, 
see Ref.~\cite{wang2010}. 
It should be noted that the calculations in Ref.~\cite{wang2010} were carried
  out with the simplified local Hamiltonian $\hat{H}^{\rm dens}_{c}$  in 
Eq.~(\ref{app3.5}). Taking the full atomic interaction into account may therefore change 
 results, at least quantitatively. Nevertheless, these results already 
illustrate how 
 important it is, 
 in studies on transition metal compounds, to treat 
 the local Coulomb interaction in a more sophisticated way then provided 
 by state-of-art DFT methods.

\section{Summary and Outlook}
In this tutorial presentation, we have provided a comprehensive 
 introduction into the Gutzwiller variational approach and its 
 merger with the density functional theory. 
 Numerically, the Gutz\-willer method is rather `cheap'  as compared 
to other many-particle approaches. It will therefore, quite likely, emerge as
   an important tool for the improvement of existing ab-initio methods.
There are two more recent developments 
which we shall briefly mention as an outlook:

\begin{itemize} 
 \item { \bf The time-dependent Gutzwiller theory} 

 The Gutzwiller theory, as introduced in this presentation, 
 can be used for the calculation of ground-state properties
 and of quasi-particle energies in the Fermi-liquid regime. 
 For the description of experiments one often needs to calculate
 two-particle response functions such as the magnetic susceptibility 
 or the optical conductivity. This is achieved by the so-called
 `time dependent Gutzwiller theory'. This method is derived 
 in a very similar way as the `random-phase approximation' can be introduced 
 as a time-dependent generalisation of the Hartree--Fock theory. 
 It was first developed 
 for single-band Hubbard models by Seibold et al.\ 
\cite{seibold1998,seibold2001}  
 and has been applied with astonishing success 
to quite a number of 
 such models and response functions \cite{seibold1998b,seibold2003,lorenzana2003,seibold2004,seibold2004b,lorenzana2005,seibold2005,seibold2006,seibold2007,seibold2008,seibold2008b}.
 Recently, the method has been generalised for the study of 
multi-band models~\cite{buenemann2011b,buenemann2011c}.

 \item { \bf Beyond the Gutzwiller approximation}

As we have demonstrated in this presentation, the energy-functional which 
we derived in infinite dimensions (i.e., the Gutzwiller approximation), 
constitutes already a major improvement
 over, e.g., the Hartree--Fock theory. It is well-known, however,
 that the limit of infinite spatial dimensions has some severe limitations.
 For example, if we consider the Fermi surface of a single-band Hubbard 
model, it will be independent of $U$ as long as no symmetry-broken 
 phases are considered. This cannot be correct in finite dimensions
 which becomes evident already from straightforward 
 perturbation theory in $U$~\cite{halboth1997}. 
It is also known from a numerical 
 evaluation of Gutzwiller wave functions in two dimensions that, 
for sufficiently large 
 values of $U$, the variational ground states can be superconducting
\cite{eichenberger2007,eichenberger2009}. 
 This is also not reproduced within the Gutzwiller approximation. 
 In a recent work, we have therefore proposed an efficient 
 diagrammatic method for the evaluation of  Gutzwiller wave functions
 in finite dimensions~\cite{buenemann2012}. It has enabled us to study 
 correlation-induced Fermi-surface deformations~\cite{buenemann2012}  
as well as 
 superconductivity (unpublished). The numerical efforts of this 
 method are still moderate and the investigation of more complicated 
  multi-band models will therefore be feasible in the near future.

\end{itemize} 

\vspace{0.2cm}
{\bf \Large Acknowledgement}
\vspace{0.2cm}

The proofreading of this manuscript by F. Gebhard is gratefully acknowledged. I also thank the authors of Ref.~\cite{wang2010} for providing me with the data, shown in Fig.~\ref{fig10b}.

\section*{Appendix}
\appendix
\section{Minimisation of functions with respect to non-interacting 
density matrices}\label{appen1}

We consider a general function $E(\tilde{\rho})$ of a non-interacting 
 density matrix $\tilde{\rho}$ with the elements
 \begin{equation}
\rho_{\gamma,\gamma'}=\langle \hcd_{\gamma'} \hc_{\gamma}\rangle_{\Phi_0}\;.
\end{equation}
The fact that  $\tilde{\rho}$ is derived from a 
 single-particle product  wave function $\ket{\Phi_0}$ is equivalent to the 
 matrix equation $\tilde{\rho}^2=\tilde{\rho}$. 
Hence, the minimum of $E(\tilde{\rho})$ in the `space' of all  
{\sl non-interacting} 
 density matrices is determined by the condition
\begin{equation}
\frac{\partial}{\partial \rho_{\gamma',\gamma}}L(\tilde{\rho})=0\;,
\end{equation}
where we introduced the `Lagrange functional'
\begin{eqnarray}\label{sfg}
L(\tilde{\rho})&\equiv& E(\tilde{\rho})-\sum_{l,m}\Omega_{l,m}
\big[\tilde{\rho}^2-\tilde{\rho}\big]_{m,l} \\
&=& E(\tilde{\rho})-\sum_{l,m}\Omega_{l,m}\Big(
\sum_p \rho_{m,p} \rho_{p,l}-\rho_{m,l}\Big)
\end{eqnarray}
and the matrix $\tilde{\Omega}$ of Lagrange parameters $\Omega_{l,m}$. 
The minimisation of~(\ref{sfg}) leads to the matrix equation
\begin{equation}\label{zuaer}
\tilde{H}=\tilde{\rho}\tilde{\Omega}+\tilde{\Omega}\tilde{\rho}-\tilde{\Omega}
\end{equation}
for the `Hamilton matrix' $\tilde{H}$ with the elements 
\begin{equation}
H_{\gamma,\gamma'}=
\frac{\partial}{\partial \rho_{\gamma',\gamma}}
 E(\tilde{\rho})\;.
\end{equation}
For density matrices which satisfy $\tilde{\rho}^2=\tilde{\rho}$, 
Eq.~(\ref{zuaer}) leads to $[\tilde{H},\tilde{\rho}]=0$.
Hence, $\tilde{H}$ and $\tilde{\rho}$ must have the same basis 
 of (single-particle) eigenvectors and, consequently, 
$\ket{\Phi_0}$ is the ground state of
 \begin{equation}
\hat{H}_0^{\rm eff}=\sum_{\gamma,\gamma'}H_{\gamma,\gamma'}
\hcd_{\gamma} \hc_{\gamma'}\;.
\end{equation}

\newpage
 \bibliographystyle{unsrt}
\bibliography{bib3}

\begin{thebibliography}{10}

\bibitem{hohenberg1964}
P.~Hohenberg and W.~Kohn.
\newblock {\em Phys.~Rev.}, 136:864, 1964.

\bibitem{sugano1970}
S.~Sugano, Y.~Tanabe, and H.~Kamimura.
\newblock {\em {Multiplets of Transition-Metal Ions in Crystals}}.
\newblock Pure and Applied Physics 33, Academic Press, {New York}, 1970.

\bibitem{feenberg1969}
E.~Feenberg.
\newblock {\em {Theory of Quantum Liquids}}.
\newblock Academic Press, {New York}, 1969.

\bibitem{clements1993}
B.~E. Clements, E.~Krotscheck, J.~A. Smith, and C.~E. Campbell.
\newblock {\em Phys.~Rev.~B}, 47:5239, 1993.

\bibitem{gutzwiller1963}
M.C. Gutzwiller.
\newblock {\em Phys.~Rev.~Lett}, 10:159, 1963.

\bibitem{gutzwiller1964}
M.C. Gutzwiller.
\newblock {\em Phys.~Rev.}, 134:A923, 1964.

\bibitem{gutzwiller1965}
M.C. Gutzwiller.
\newblock {\em Phys.~Rev.}, 137:A1726, 1965.

\bibitem{buenemann1998}
J.~B{\"u}nemann, W.~Weber, and F.~Gebhard.
\newblock {\em Phys.~Rev.~B}, 57:6896, 1998.

\bibitem{buenemann2005}
J.~B{\"u}nemann, F.~Gebhard, and W.~Weber.
\newblock In A.~Narlikar, editor, {\em Frontiers in Magnetic Materials}.
  Springer, Berlin, 2005.

\bibitem{buenemann2012b}
J.~B{\"u}nemann, T.~Schickling, F.~Gebhard, and W.~Weber.
\newblock {\em physica status solidi (b)}, 249:1282, 2012.

\bibitem{buenemann2005b}
J.~B{\"u}nemann, K.~J\'avorne-Radn\'oczi, P.~Fazekas, and F.~Gebhard.
\newblock {\em J. Phys. Cond. Matt}, 17:3807, 2005.

\bibitem{metzner1988}
W.~Metzner and D.~Vollhardt.
\newblock {\em Phys.~Rev.~B}, 37:7382, 1988.

\bibitem{kollar2002}
M.~Kollar and D.~Vollhardt.
\newblock {\em Phys.~Rev.~B}, 65:155121, 2002.

\bibitem{gebhard1987}
F.~Gebhard and D.~Vollhardt.
\newblock {\em Phys.~Rev.~Lett.}, 59:1472, 1987.

\bibitem{gebhard1988}
F.~Gebhard and D.~Vollhardt.
\newblock {\em Phys.~Rev.~B}, 38:6911, 1988.

\bibitem{gebhard1990}
F.~Gebhard.
\newblock {\em Phys.~Rev.~B}, 41:9452, 1990.

\bibitem{buenemann1998b}
J.~B{\"u}nemann.
\newblock {\em Eur.~Phys.~J.~B}, 4:29, 1998.

\bibitem{note2}
Note that, in general, the fields $\eta_{\sigma}$ are non-diagonal
  ($\eta_{\sigma,\sigma'}$) and given as the derivative of the
  energy-functional with respect to the elements of the non-interacting density
  matrix~(\ref{8.12a}). This requires the evaluation of the more complicated
  energy functional without the condition~(\ref{8.12}), see
  Ref.\cite{buenemann2012b}.

\bibitem{buenemann2003b}
J.~B{\"u}nemann, F.~Gebhard, and R.~Thul.
\newblock {\em Phys.~Rev.~B}, 67:75103, 2003.

\bibitem{buenemann1997b}
J.~B{\"u}nemann and W.~Weber.
\newblock {\em Physica~B}, 230:412, 1997.

\bibitem{cruz2008}
C.~de~la Cruz, Q.~Huang, J.~W. Lynn, J.~Li, W.~Ratcliff, J.~L. Zarestky, H.~A.
  Mook, G.~F. Chen, J.~L. Luo, N.~L. Wang, and P.~Dai.
\newblock {\em Nature~London}, 453:899, 2008.

\bibitem{qureshi2010}
N.~Qureshi, Y.~Drees, J.~Werner, S.~Wurmehl, C.~Hess, R.~Klingeler,
  B.~B{\"u}chner, M.~T. Fern{\'a}ndez-D{\'\i}az, and M.~Braden.
\newblock {\em Phys.~Rev.~B}, 82:184521, 2010.

\bibitem{li2010}
H.~F. Li, H.-F. Li, W.~Tian, J.-Q. Yan, J.~L. Zarestky, R.~W. McCallum, T.~A.
  Lograsso, and D.~Vaknin.
\newblock {\em Phys.~Rev.~B}, 82:064409, 2010.

\bibitem{mazin2008}
E.~I. Mazin, M.~D. Johannes, L.~Boeri, K.~Koepernik, and D.~J. Singh.
\newblock {\em Phys.~Rev.~B}, 78:085104, 2008.

\bibitem{zhang2010}
Y.-Z. Zhang, I.~Opahle, H.~O. Jeschke, and R.~Valent{\'i}.
\newblock {\em Phys.~Rev.~B}, 81:094505, 2010.

\bibitem{andersen2010}
O.~K. Andersen and L.~Boeri.
\newblock {\em Ann. Physik (Berlin)}, 523:8, 2011.

\bibitem{graser2009}
S.~Graser, T.~Maier, P.~Hirschfeld, and D.~Scalapino.
\newblock {\em New Journal of Physics}, 11:025016, 2009.

\bibitem{zhou2010}
S.~Zhou and Z.~Wang.
\newblock {\em Phys. Rev. Lett.}, 105:096401, 2010.

\bibitem{buenemann2011}
T.~Schickling, F.~Gebhard, and J.~B{\"u}nemann.
\newblock {\em Phys. Rev. Lett.}, 106:146402, 2011.

\bibitem{ishida2010}
H.~Ishida and A.~Liebsch.
\newblock {\em Phys.~Rev.~B}, 81:054513, 2010.

\bibitem{ho2008}
K.~M. Ho, J.~Schmalian, and C.~Z. Wang.
\newblock {\em Phys.~Rev.~B}, 77:073101, 2008.

\bibitem{deng2008}
X.~Deng, X.~Dai, and Z.~Fang.
\newblock {\em Europhys. Lett.}, 83:37008, 2008.

\bibitem{deng2009}
X.~Deng, L.~Wang, X.~Dai, and Z.~Fang.
\newblock {\em Phys.~Rev.~B}, 79:075114, 2009.

\bibitem{wang2008}
G.~Wang, X.~Dai, and Z.~Fang.
\newblock {\em Phys. Rev. Lett.}, 101:066403, 2008.

\bibitem{wang2010}
G.~Wang, Y.~M. Qian, G.~Xu, X.~Dai, and Z.~Fang.
\newblock {\em Phys. Rev. Lett.}, 104:047002, 2010.

\bibitem{weng2011}
H.~Weng, G.~Xu, H.~Zhang, S.-C.\ Zhang, X.~Dai, and Z.~Fang.
\newblock {\em Phys.~Rev.~B}, 84:060408, 2011.

\bibitem{karolak2010}
M.~Karolak, G.~Ulm, T.~Wehling, V.~Mazurenko, A.~Poteryaev, and
  A.~Lichtenstein.
\newblock {\em Journal of Electron Spectroscopy and Related Phenomena}, 181:11,
  2010.

\bibitem{seibold1998}
G.~Seibold, E.~Sigmund, and V.~Hizhnyakov.
\newblock {\em Phys.~Rev.~B}, 57:6937, 1998.

\bibitem{seibold2001}
G.~Seibold and J.~Lorenzana.
\newblock {\em Phys.~Rev.~Lett.}, 86:2605, 2001.

\bibitem{seibold1998b}
G.~Seibold.
\newblock {\em Phys.~Rev.~B}, 58:15520, 1998.

\bibitem{seibold2003}
G.~Seibold, F.~Becca, and J.~Lorenzana.
\newblock {\em Phys.~Rev.~B}, 67:085108, 2003.

\bibitem{lorenzana2003}
J.~Lorenzana and G.~Seibold.
\newblock {\em Phys.~Rev.~Lett.}, 90:066404, 2003.

\bibitem{seibold2004}
G.~Seibold and J.~Lorenzana.
\newblock {\em Phys.~Rev.~B}, 69:134513, 2004.

\bibitem{seibold2004b}
G.~Seibold, F.~Becca, P.~Rubin, and J.~Lorenzana.
\newblock {\em Phys.~Rev.~B}, 69:155113, 2004.

\bibitem{lorenzana2005}
J.~Lorenzana, G.~Seibold, and R.~Coldea.
\newblock {\em Phys.~Rev.~B}, 72:224511, 2005.

\bibitem{seibold2005}
G.~Seibold and J.~Lorenzana.
\newblock {\em Phys.~Rev.~Lett.}, 94:107006, 2005.

\bibitem{seibold2006}
G.~Seibold and J.~Lorenzana.
\newblock {\em Phys.~Rev.~B}, 73:144515, 2006.

\bibitem{seibold2007}
G.~Seibold and J.~Lorenzana.
\newblock {\em Journal of Superconductivity and Novel Magnetism}, 20:619, 2007.

\bibitem{seibold2008}
G.~Seibold, F.~Becca, and J.~Lorenzana.
\newblock {\em Phys.~Rev.~Lett.}, 100:016405, 2008.

\bibitem{seibold2008b}
G.~Seibold, F.~Becca, and J.~Lorenzana.
\newblock {\em Phys.~Rev.~B}, 78:045114, 2008.

\bibitem{buenemann2011b}
E.~v.~Oelsen, G.~Seibold, and J.~B{\"u}nemann.
\newblock {\em Phys.~Rev.~Lett.}, 107:076402, 2011.

\bibitem{buenemann2011c}
E.~v.~Oelsen, G.~Seibold, and J.~B{\"u}nemann.
\newblock {\em New J.~Phys.}, 13:113031, 2011.

\bibitem{halboth1997}
C.~J. Halboth and W.~Metzner.
\newblock {\em Z.~Phys.~B}, 102:501, 1997.

\bibitem{eichenberger2007}
D.~Eichenberger and D.~Baeriswyl.
\newblock {\em Phys.~Rev.~B}, 76:180504, 2007.

\bibitem{eichenberger2009}
D.~Eichenberger and D.~Baeriswyl.
\newblock {\em Phys.~Rev.~B}, 79:100510, 2009.

\bibitem{buenemann2012}
J.~B{\"u}nemann, T.~Schickling, and F.~Gebhard.
\newblock {\em Europhys. Lett.}, 98:27006, 2012.

\end{thebibliography}
\end{document}